\documentclass[final,1p,times]{elsarticle}
\biboptions{sort&compress}

\usepackage{amssymb}
\usepackage{amsmath}

\usepackage{hyperref} 
\usepackage{booktabs} 
\usepackage{fvextra}  
\usepackage[table]{xcolor} 
\usepackage{inconsolata} 
\usepackage{tikz}
\usepackage{rotating} 
\usepackage{placeins} 
\usepackage{textpos}

\usetikzlibrary{shapes.geometric, arrows.meta, plotmarks, positioning}

\definecolor{good}{HTML}{006600}
\definecolor{bad}{HTML}{C00000}
\newcommand{\llmonline}{\textcolor{good}{Online}}
\newcommand{\llmoffline}{\textcolor{bad}{Offline}}

\makeatletter
\renewcommand{\ps@pprintTitle}{%
  \let\@oddhead\@empty
  \let\@evenhead\@empty
  \def\@oddfoot{\reset@font\hfil\thepage\hfil}
  \let\@evenfoot\@oddfoot
}
\makeatother

\newcommand{\drawHistogram}[5]{

    \pgfmathsetmacro{\sumValues}{#1 + #2 + #3 + #4 + #5}

    \pgfmathsetmacro{\scaleFactor}{2}
    
    \pgfmathsetmacro{\normA}{(#1 / \sumValues) * \scaleFactor}
    \pgfmathsetmacro{\normB}{(#2 / \sumValues) * \scaleFactor}
    \pgfmathsetmacro{\normC}{(#3 / \sumValues) * \scaleFactor}
    \pgfmathsetmacro{\normD}{(#4 / \sumValues) * \scaleFactor}
    \pgfmathsetmacro{\normE}{(#5 / \sumValues) * \scaleFactor}

    \begin{tikzpicture}
        \path (-0.25,-0.25) (5.25,1.75 * \scaleFactor);
        \draw[gray,line width=4.5pt] plot coordinates { 
            (0,0) 
            (0, \normA) 
            (1, \normA) 
            (1, \normB) 
            (2, \normB) 
            (2, \normC) 
            (3, \normC) 
            (3, \normD) 
            (4, \normD) 
            (4, \normE) 
            (5, \normE)
            (5, 0)  
        };
        \foreach \i in {0,...,5} {
            \path plot[mark=square*,mark options={color=black},mark size=1.5pt] coordinates { (\i, 0) };
        }
        
    \end{tikzpicture}
}

\makeatletter
\def\ps@pprintTitle{   \let\@oddhead\@empty
   \let\@evenhead\@empty
   \def\@oddfoot{\reset@font\hfil\thepage\hfil}
   \let\@evenfoot\@oddfoot
}
\makeatother

\journal{arxiv}

\begin{document}

\begin{frontmatter}

\title{GPT-4.1 Sets the Standard in Automated Experiment Design Using Novel Python Libraries} 

\author[label1,label2]{Nuno Fachada}
\author[label1]{Daniel Fernandes}
\author[label1,label2]{Carlos M. Fernandes}
\author[label1,label3]{Bruno D. Ferreira-Saraiva}
\author[label2,label4]{João P. Matos-Carvalho}

\affiliation[label1]{organization={Copelabs, Lusófona University},
            addressline={Campo Grande, 376},
            city={Lisboa},
            postcode={1749-024},
            country={Portugal}}

\affiliation[label2]{organization={Center of Technology and Systems (UNINOVA-CTS) and Associated Lab of Intelligent Systems (LASI)},
            city={Caparica},
            postcode={2829-516},
            country={Portugal}}

\affiliation[label3]{organization={CICANT, Lusófona University},
            addressline={Campo Grande, 376},
            city={Lisboa},
            postcode={1749-024},
            country={Portugal}}

\affiliation[label4]{organization={LASIGE, Departamento de Informática, Faculdade de Ciências, Universidade de Lisboa, 1749--016 Lisboa, Portugal},
}

\begin{abstract}
Large Language Models (LLMs) have advanced rapidly as tools for automating code generation in scientific research, yet their ability to interpret and use unfamiliar Python APIs for complex computational experiments remains poorly characterized. This study systematically benchmarks a selection of state-of-the-art LLMs in generating functional Python code for two increasingly challenging scenarios: conversational data analysis with the \textit{ParShift} library, and synthetic data generation and clustering using \textit{pyclugen} and \textit{scikit-learn}. Both experiments use structured, zero-shot prompts specifying detailed requirements but omitting in-context examples. Model outputs are evaluated quantitatively for functional correctness and prompt compliance over multiple runs, and qualitatively by analyzing the errors produced when code execution fails.
Results show that only a small subset of models consistently generate correct, executable code. GPT-4.1 achieved a 100\% success rate across all runs in both experimental tasks, whereas most other models succeeded in fewer than half of the runs, with only Grok-3 and Mistral-Large approaching comparable performance.
In addition to benchmarking LLM performance, this approach helps identify shortcomings in third-party libraries, such as unclear documentation or obscure implementation bugs. Overall, these findings highlight current limitations of LLMs for end-to-end scientific automation and emphasize the need for careful prompt design, comprehensive library documentation, and continued advances in language model capabilities.
\end{abstract}


\end{frontmatter}

\begin{textblock*}{190mm}(-3cm,-20.18cm)
    {\noindent \footnotesize \color{black!90} The peer-reviewed version of this paper is
    published in Future Internet at \url{https://doi.org/10.3390/fi17090412}.
    This version is typeset by the authors and differs only in pagination and
    typographical detail.}
\end{textblock*}

\section{Introduction}

The automation of computational experiments is a critical component of modern scientific research, enabling efficient exploration of hypotheses, reproducible analysis pipelines, and scalable data-driven discovery. In recent years, large language models (LLMs) have emerged as powerful tools capable of generating executable code from natural language descriptions~\cite{chen2021evaluating,10.1145/3510003.3510203,gu2024effectiveness,fernandes2025deepseek}, raising the possibility of automating significant portions of the experimental workflow with minimal human intervention~\cite{odonoghue2023bioplannerautomaticevaluationllms,hong2024data,charness2025next}. These advancements hold particular promise for researchers and practitioners who wish to rapidly prototype, modify, or extend computational analyses without requiring deep expertise in software engineering or specific programming libraries.

Despite rapid progress in LLM capabilities, a central challenge remains: the reliable generation of correct and functional code in larger computational experiments, especially when tasks involve lesser-known third-party libraries with limited online presence beyond their official documentation. The ability of LLMs to parse, interpret, and accurately use unfamiliar application programming interfaces (APIs) is crucial for their broader adoption as assistants or agents in real-world scientific settings. Current evidence suggests that LLM performance varies widely with prompt design, library familiarity, and task complexity, often requiring expert oversight or extensive validation to ensure robustness and correctness~\cite{amatriain2024prompt,chen2021evaluating,gu2024effectiveness,spiess2025calibration}.

This study systematically evaluates the ability of state-of-the-art LLMs to generate functional Python code for complex computational experiments involving two distinct and progressively challenging tasks. Both tasks require the correct integration of specialized Python libraries---\textit{ParShift}~\cite{ferreira2023parshift} for conversational data analysis and \textit{pyclugen}~\cite{fachada2023generating} for synthetic data generation---posing realistic scenarios where knowledge of obscure APIs is essential. The evaluation considers a diverse set of leading LLMs, each prompted in a zero-shot setting to maximize reproducibility and emulate the experience of a domain expert querying the model with a technically precise request~\cite{amatriain2024prompt,10.5555/3600270.3602070}.

To achieve a comprehensive assessment, the study employs a multi-level analytical framework. Model outputs are scored for correctness and functional compliance; performance is quantified in terms of overall success rates, and the success rates of different models are statistically compared to determine whether observed differences are significant. Additionally, the variability and consistency of generated outputs are systematically analyzed, providing insight into the stochastic nature of LLM-generated code and the factors influencing its reliability. When outputs fail to execute correctly, error messages and failure modes are qualitatively examined to identify recurring issues and diagnose challenges related to API misuse, type mismatches, and prompt misinterpretation. Finally, for the subset of fully correct outputs, code quality is examined through several software engineering metrics, summarized using descriptive statistics and complemented by an analysis of distributional characteristics to provide a quantitative assessment.

By rigorously benchmarking LLMs on realistic, domain-specific tasks, this work helps characterize the current capabilities of automated code generation for computational experiments. The results show that only a small subset of models---GPT-4.1, Grok 3, and Mistral Large---consistently produced correct, executable code throughout the evaluated tasks. Beyond model performance, the study reinforces the relevance of structured zero-shot prompts as a practical and reproducible baseline for evaluating LLMs in isolation, prior to the confounding effects of multi-shot examples, self-refinement, or external tool use. Moreover, the study demonstrates the potential that these types of experiments have in identifying documentation gaps and edge cases in third-party libraries, offering valuable feedback for library authors and developers of scientific software.

The subsequent sections of this paper are organized as follows. Section~\ref{sec:background} provides background information on the use of LLMs for domain-specific code generation and the automation of computational experiments. Section~\ref{sec:methods} describes the materials and methods employed, including the problem context, prompt design, model selection, experimental setup, and evaluation procedures. Results are presented in Section~\ref{sec:results}, followed by a detailed discussion in Section~\ref{sec:discussion}. Section~\ref{sec:limitations} outlines the study's limitations and opportunities for future research. Finally, Section~\ref{sec:conclusions} concludes the paper with final remarks and recommendations.

\section{Background}
\label{sec:background}

This section reviews relevant literature and provides essential context on how LLMs are being used to support scientific research. It begins by describing a range of scientific tasks where LLMs have already proven useful, such as forming hypotheses, conducting peer-review and performing literature review. Then, it examines frameworks that convert natural-language protocols into executable workflows, manage experimental variables and constraints, and generate ready-to-run code. The final subsection highlights advances in crafting effective prompts to steer model behavior and ensure consistent and reproducible outputs in scientific experimentation.

\subsection{LLM in Research and Development}

The integration of LLMs into scientific research introduces new workflows in research and development, wherein natural language processing systems assist or autonomously execute a range of scientific tasks, including protocol optimization, data analysis, and experimental design.

A key application of LLMs is automating hypothesis generation. In~\cite{xiong2024improvingscientifichypothesisgeneration}, for instance, Xiong et al. introduce KG-CoI, a system that integrates external structured and unstructured domain-specific knowledge from scientific resources into LLMs, including hallucination detection, to support more reliable scientific exploration. KG-CoI significantly reduces factual errors and increases the relevance of generated hypotheses. In~\cite{pu2025piflowprincipleawarescientificdiscovery}, the authors propose PiFlow, a multi-agent LLM framework that integrates domain-specific scientific principles into hypothesis generation to overcome LLMs' difficulty in consistently linking hypotheses with evidence. PiFlow achieved a 94\% improvement in solution quality compared to a single LLM conducting the discovery process independently, across applications in nanomaterials and biomolecular research.

Jin et al.~\cite{jin2024agentreviewexploringpeerreview} present AgentReview, a simulation framework where LLMs emulate the roles of reviewers, authors, and area chairs to analyze biases and decision variability in scientific peer review. By modeling latent sociological factors like authority bias and altruism fatigue, the study reveals that LLM-based agents can offer insights into improving the fairness and robustness of scientific evaluation systems.

In~\cite{agarwal2025litllmtoolkitscientificliterature}, Agarwal et al. introduce LitLLM, a system for generating literature reviews from an abstract using off-the-shelf LLMs. The authors show that LitLLM substantially reduces time and effort for literature review compared to traditional methods.

As the studies reviewed above attest, LLMs are already reshaping key phases of the scientific workflow. However, translating ideas into experiments still depends on detailed protocol planning and execution. In the following section, we discuss how LLM-based agents can draft detailed experimental protocols, and adapt workflows in real time, effectively turning high-level research designs into implementable laboratory action.

\subsection{LLMs for Experiment Planning and Implementation}

LLMs are increasingly being integrated into computational experiment design workflows, particularly through Python and its rich set of libraries. These pipelines automate repetitive tasks, shorten iteration times, and improve reproducibility~\cite{luo2025llm4srsurveylargelanguage}. Recently, Charness et al.~\cite{charness2025next} argued that LLMs can or at least could soon be integrated into the design, implementation and analysis of social and economic experiments. The remainder of this subsection examines representative approaches to harnessing these capabilities for robust and scalable experiment planning and implementation.

One of the already established abilities of LLMs in assisting scientific research is code generation, either by producing ready-to-run code or by supporting researchers in the iterative development of scripts, pipelines, and analytical workflows. Over time, LLMs have moved beyond simple code snippets to domain-specific pipelines, enabling more sophisticated prototyping and documentation.

In~\cite{10160591}, the authors adapt LLMs trained on code completion for writing robot policy code according to natural language prompts. The generated robot policies exhibit spatial-geometric reasoning and are able to prescribe precise values to ambiguous descriptions. By relying on a hierarchical prompting strategy, their approach is able to write more complex code and solve 39.8\% of the problems on the HumanEval~\cite{chen2021evaluating} benchmark.

Luo et al.~\cite{LUO2024100488} use LLMs to generate robot control programs, testing and optimizing the output in a simulation environment. After a number of optimization rounds, the robot control codes are deployed on a real robot for construction assembly tasks. The experiments show that their approach can improve the quality of the generated code, thus simplifying the robot control process and facilitating the automation of construction tasks.

In~\cite{fernandes2025deepseek}, the authors present a comparative evaluation of 16 LLMs in generating Python code for UAV placement and signal power calculations in LoRaWAN environments, demonstrating that state-of-the-art models like DeepSeek-V3 and GPT-4 consistently produce accurate solutions, while smaller, locally executable models such as Phi-4 and LLaMA-3.3 also show strong performance, highlighting the viability of lightweight alternatives. The study demonstrates the importance of domain-specific fine-tuning, offering valuable insights into the practical deployment of LLMs for specialized engineering tasks.

LLMs can write individual scripts, but full experimental design demands additional abstraction, planning, and domain expertise. It requires integrating hypotheses, protocol structure, data collection strategies, ethical considerations, and analysis plans, a complex orchestration far beyond simple script generation.

A method for evaluating LLMs' ability to generate and interpret experimental protocols is proposed in~\cite{odonoghue2023bioplannerautomaticevaluationllms}. The authors describe BioPlanner, a framework that converts natural language protocols into pseudocode and then assesses a model’s ability to reconstruct those protocols using a predefined set of admissible pseudofunctions. This approach enables robust, automated evaluation of long-horizon planning tasks in biology, reducing reliance on manual review and paving the way for scalable scientific automation.

Hong et al.~\cite{hong2024data} introduce Data Interpreter, an LLM-based agent designed to automate end-to-end data science workflows, addressing the limitations of existing LLMs in handling complex and dynamic data science tasks. The system decomposes complex data science tasks into graph-based subproblems and dynamically adapts workflows via programmable node generation. It embeds a confidence-driven verifier to flag and correct logical inconsistencies in generated code. In benchmarks like InfiAgent-DABench and MATH, it achieves roughly 25–26\% gains over open source baselines.

This is where our proposal takes shape: it evaluates how well an LLM can parse APIs and assemble complete computational experiments from a single prompt, without further human guidance. While this setup is not agentic in nature, it does bear a superficial resemblance to recent work on LLM-based agents, such as Hong et al.'s Data Interpreter~\cite{hong2024data}, ReAct~\cite{yao2023reactsynergizingreasoningacting}, or Auto-GPT~\cite{yang2023autogptonlinedecisionmaking}. These frameworks emphasize multi-step reasoning, interactive planning, and iterative tool use, often allowing models to execute, verify, and refine their own outputs. In contrast, the methodology used in this paper deliberately isolates and benchmarks a model's intrinsic ability to interpret unfamiliar APIs and return fully functional code in a single pass, guided only by a zero-shot structured prompt. In this sense, our study is complementary to agent research: by establishing a baseline of core code generation competence, it provides a foundation for understanding the prerequisites upon which more complex agentic systems ultimately depend. Prompt design thus becomes pivotal to the scope of this investigation, and accordingly, the next section focuses on prompt engineering and its foundational principles.

\subsection{Prompt Design}
\label{sec:prompt_design}

The piece of text or set of instructions that a user provides to an LLM to elicit a specific response is referred to as a prompt. Designing effective prompts has therefore become essential to harness the full capabilities of LLMs, and in recent years this craft has evolved into a distinct field of research and development~\cite{amatriain2024prompt}. Advanced prompting techniques---such as in-context learning \cite{brown2020language}, chain-of-thought~\cite{10.5555/3600270.3602070, LUO2024100488}, and ReAct \cite{yao2023reactsynergizingreasoningacting}---are frequently employed in studies to improve the reliability and precision of experimental planning within LLM-assisted workflows.

Prompting strategies are often described along multiple dimensions. One common distinction is between structured and unstructured prompts. Structured prompts use precise formulations, with clearly defined inputs, outputs, and constraints, and tend to yield more consistent results---particularly in tasks like code generation. However, structured prompts often demand detailed knowledge of both the problem space and the model's behavior, which can hinder accessibility for non-experts. Unstructured prompting, in contrast, adopts a more conversational style, making it easier to use in real-world scenarios where users might lack technical expertise. Yet this flexibility comes at a cost, as it can lead to output inconsistencies due to the inherent ambiguity of informal language. Table~\ref{tab:prompttypes} illustrates a minimal example of this structured vs. unstructured prompting dimension.

\begin{table}[tb]
\centering
\caption{Minimal example of unstructured and structured prompts for the same programming task. The unstructured prompt is informal and open-ended, while the structured prompt specifies function details and expected input/output, illustrating the difference in detail and reproducibility.}\label{tab:prompttypes}
\begin{tabular}{rl}
\toprule
\textbf{Prompt Type} & \textbf{Prompt Example} \\
\midrule
Unstructured &
\begin{minipage}[t]{10.5cm}
\begin{Verbatim}[fontsize=\footnotesize,breaklines,breaksymbolleft={}, breaksymbolright={}, breakindent=0pt]
How do I sum a list of whole numbers in Python?
\end{Verbatim}
\end{minipage}
\\ 
Structured &
\begin{minipage}[t]{10.5cm}
\begin{Verbatim}[fontsize=\footnotesize,breaklines,breaksymbolleft={}, breaksymbolright={}, breakindent=0pt]
Write a Python function `sum_numbers(numbers: list[int]) -> int` that returns the sum of a list of integers.
\end{Verbatim}
\end{minipage}
\\
\bottomrule
\end{tabular}
\end{table}

Prompts can also be distinguished by the number of examples they provide to guide the model: zero-shot prompts offer no examples, one-shot prompts include a single instance, and few-shot prompts present several illustrations. Each style carries its own trade-offs, as supported by empirical findings---for instance, Liang et al.~\cite{10160591} showed that structured prompts containing code consistently outperform those written in plain language for tasks involving robotic reasoning. That said, ongoing advancements in LLM capabilities continue to narrow this gap, with recent studies demonstrating promising results for natural-language-based prompting in specialized domains like robotics~\cite{vemprala2023chatgpt}.

Complementing existing techniques, our method employs a fully structured, zero-shot prompt that embeds the relevant API docstrings and specifies, without examples, every step needed to assemble a complete computational experiment. This design tests the model's ability to interpret unfamiliar Python APIs, assemble complete computational workflows---from data extraction and analysis to multi-step data generation, clustering, and evaluation---and return fully functional solutions in a single pass.

\section{Materials and Methods}
\label{sec:methods}

This section details the methodology adopted to assess the ability of state-of-the-art LLMs to interpret and use Python APIs for automated computational experiment design. The following subsections describe the experimental scenarios and prompt construction (\ref{sec:methods:scenarios}), evaluation pipeline (\ref{sec:methods:impl}), selection of LLMs (\ref{sec:methods:llms}), experimental protocol (\ref{sec:methods:expsetup}), and data analysis approach (\ref{sec:methods:datan}).

\subsection{Scenarios and Prompts}
\label{sec:methods:scenarios}

To evaluate the performance of the selected LLMs (discussed in Section~\ref{sec:methods:llms}), two experimental scenarios were prepared, each associated with a carefully designed, structured zero-shot prompt. The zero-shot approach, which does not provide examples within the prompt, was chosen to maximize reproducibility, minimize potential bias from hand-crafted demonstrations, and more closely reflects typical initial queries issued by domain experts to LLMs for technical tasks~\cite{amatriain2024prompt,10.5555/3600270.3602070}. Zero-shot prompting also facilitates direct benchmarking of models' intrinsic capabilities in code generation and comprehension.

Both prompts employ a highly structured format, specifying requirements such as the function name, input and output types, the set of allowable libraries, coding standards (e.g., indentation and style), and the necessity for the function to be self-contained. This structure serves to minimize ambiguity, streamline post-processing, and ensure fair and consistent evaluation between models and runs~\cite{amatriain2024prompt,10.5555/3600270.3602070}. Moreover, by explicitly detailing the APIs to use, including relevant docstrings, the prompts challenge models to demonstrate genuine understanding and integration of unfamiliar libraries, rather than relying on memorized patterns from training data. However, since the documentation of the tested libraries is publicly available since 2023, a degree of prior exposure cannot be fully ruled out. 

The two experimental scenarios were designed to reflect progressively increasing complexity and to probe distinct aspects of LLM-driven code synthesis using recent, peer-reviewed research software. Both the \textit{ParShift}~\cite{ferreira2023parshift} and \textit{pyclugen}~\cite{fachada2023generating} libraries, although not widely known, are open source, fully unit tested, extensively documented, and readily available via the Python Package Index, thus providing a reliable and transparent foundation for the study.

\begin{table}[!htb]
    
    \caption{Prompt used in Experiment~1 (\textit{ParShift}). The table shows the full prompt excluding API docstrings; the complete prompt, including all referenced docstrings, is available in the supplementary material~\cite{fachada2025suppl}.}

    \label{tab:prompt_parshift}
    \centering
    \begin{tabular}{l}

        \toprule
        \small{Prompt 1}\\
        \midrule

\begin{minipage}{0.97\linewidth}
\begin{Verbatim}[fontsize=\scriptsize,breaklines,breaksymbolleft={}, breaksymbolright={}, breakindent=0pt]
Create a Python function named `do_parshift_exp()` that analyses CSV files containing conversations as described below. This function accepts a list of file paths (`List[str]`), each pointing to a CSV file, and returns a corresponding list of floats.

Specifically, the analysis of these conversations should be done with the **parshift** library, whose API documentation is provided below in the form the respective source docstrings:

```python
{{DOCSTRINGS HERE}}
```

The `do_parshift_exp()` function must analyse the conversations in the CSV files as follows:

1. For each CSV file specified in the list given as the function's argument, extract the proportion of the "Turn Usurping" class (as a float). Here, proportion is the sum of the "Frequency" divided by all frequencies.
2. Collect all the proportions of 'Turn Usurping' into a list of floats.
3. Return the list of floats, with one entry per file.

The `do_parshift_exp()` function should strictly follows these requirements:

- The function must be self-contained. In other words, all variables, constants, and/or helper functions must be defined within the `do_parshift_exp()` function.
- Beyond the Python standard library, only the **parshift**, **pandas**, and **numpy** libraries are allowed.
- The function should be compatible with Python 3.9 and above and follow the PEP 8 style guidelines, with 4-space indentation.
- Each input file uses a semicolon (;) as a delimiter and contains conversational data as expected by the parshift library.
- Do not include print statements, plots, or additional output.

Please respond with a single code block only. Inline comments within the code are fine, but do not include any explanation or text outside the code block.
\end{Verbatim}
\end{minipage} \\

    \bottomrule
    \end{tabular}
\end{table}

The first experiment, using \textit{ParShift}, requires LLMs to generate a Python function, \linebreak \texttt{do\_parshift\_exp()}, that processes a list of CSV files containing conversational data, extracting the proportion of utterances pairs labeled as ``Turn Usurping'' from each file, and returning the results as a list of floats. This scenario is hypothesized to be relatively straightforward, involving a linear workflow that primarily tests basic API usage, data parsing, and compliance with explicit prompt constraints. The prompt, with API docstrings omitted for space considerations, is detailed in Table~\ref{tab:prompt_parshift}.

The second experiment employs \textit{pyclugen} to significantly increase task complexity. Here, LLMs must generate a function, \texttt{do\_clustering\_exp()}, that conducts a full clustering analysis by generating multiple synthetic datasets in three dimensions (with varying cluster dimensions along a general direction), applying several clustering algorithms available in \textit{scikit-learn}~\cite{pedregosa2011scikit}, evaluating clustering quality using the $V$-measure~\cite{rosenberg2007v}, and collating detailed results (including algorithm, parameter values, and seed) in a \textit{Pandas} data frame. This multi-step pipeline tests the LLMs' ability to integrate unfamiliar APIs, manage parameterization, and synthesize robust, structured code. Potential error points include misuse of APIs (both from \textit{pyclugen} and \textit{scikit-learn}), mishandling of \textit{pyclugen} output, and inconsistent output formatting. The prompt is provided in Table~\ref{tab:prompt_pyclugen}---the API docstrings are again omitted for brevity.

\begin{table}[!htbp]
    \caption{Prompt used in Experiment~2 (\textit{pyclugen}). The table presents the prompt text without the included API docstrings; the complete version with all API docstrings can be found in the supplementary material~\cite{fachada2025suppl}.}

    \label{tab:prompt_pyclugen}
    \centering
    \begin{tabular}{l}

        \toprule
        \small{Prompt 2}\\
        \midrule

\begin{minipage}{0.97\linewidth}
\begin{Verbatim}[fontsize=\scriptsize,breaklines,breaksymbolleft={}, breaksymbolright={}, breakindent=0pt]
Create a Python function named `do_clustering_exp()` that runs a comprehensive clustering experiment. This function accepts a single `n_samples` argument of type `int`, and returns a pandas DataFrame.

Specifically, the experiment involves generating synthetic clustered datasets using the `clugen` function from the **pyclugen** library, whose API documentation is provided below in the form the respective source docstrings:

```python
{{DOCSTRINGS HERE}}
```

The computational experiment requirements are as follows:

- The dimensionality (`num_dims`) should be set to 3.
- The number of clusters (`num_clusters`) should be 10.
- Each dataset should contain exactly 10,000 points (`num_points`).
- The average cluster direction vector (`direction`) is set to a 3-dimensional vector of ones.
- The standard deviation of cluster direction angles (`angle_disp`) should be pi/16 radians.
- The average separation between cluster centers (`cluster_sep`) should be fixed at 50 for each dimension.

Additionally, perform a parameter sweep over different values of the average cluster-supporting line length (`llength`):

- The average lengths (`llength`) should range from 0 to 800, inclusive, with steps of 25 (i.e., `[0, 25, 50, ..., 800]`).

Finally:

- The length dispersion of cluster-supporting lines (`llength_disp`) should be set to 0.2 * llength.
- The cluster lateral dispersion (`lateral_disp`) should be set to 0.5 * llength.

For each dataset generated, apply the following clustering algorithms from scikit-learn:

- K-means++ (`KMeans` with `init='k-means++'`)
- Gaussian Mixture Model clustering (`GaussianMixture`)
- Agglomerative Hierarchical Clustering (`AgglomerativeClustering`) with the following linkage methods:
  - Ward linkage
  - Single linkage
  - Complete linkage
  - Average linkage (with Euclidean distance)

Please note that in recent version of scikit-learn, the `AgglomerativeClustering` class no longer accepts the `affinity` parameter. This parameter is now called `metric` which is set to `euclidean` by default.

Evaluate the clustering quality for each run using the V-measure (`v_measure_score`) provided by scikit-learn, comparing predicted cluster labels to the ground truth labels provided by `clugen`.

For every run, record the following into a pandas DataFrame:

- `run`: Sample run index (from 0 to `n_samples` - 1)
- `algorithm`: Name of the clustering algorithm (e.g., `"kmeans++"`, `"em"`, `"ahc-ward"`, `"ahc-single"`, `"ahc-complete"`, `"ahc-avg"`)
- `length`: The current average line length (`llength`)
- `vmeas`: Calculated V-measure score
- `seed`: Random seed used for generating the dataset

The `do_clustering_exp()` function should return this pandas DataFrame.

The `do_clustering_exp()` function should strictly follows these requirements:

- The function must be self-contained. In other words, all variables, constants, and/or helper functions must be defined within the `do_clustering_exp()` function.
- Beyond the Python standard library, only the **pyclugen**, **numpy**, **pandas**, **sklearn**, and **scipy** libraries are allowed.
- The function should be compatible with Python 3.9 and above and follow the PEP 8 style guidelines, with 4-space indentation.
- Do not include print statements, plots, or additional output.

Please respond with a single code block only. Inline comments within the code are fine, but do not include any explanation or text outside the code block.
\end{Verbatim}
\end{minipage} \\

    \bottomrule
    \end{tabular}
\end{table}

The two scenarios were first implemented and validated by the authors to establish the baseline results used for comparison with the LLM-generated functions. Both the baseline implementations and their results are provided in the supplementary material~\cite{fachada2025suppl}.

\subsection{Evaluation Pipeline}
\label{sec:methods:impl}

\begin{figure}[tb]
    \centering
    \includegraphics[width=0.99\linewidth]{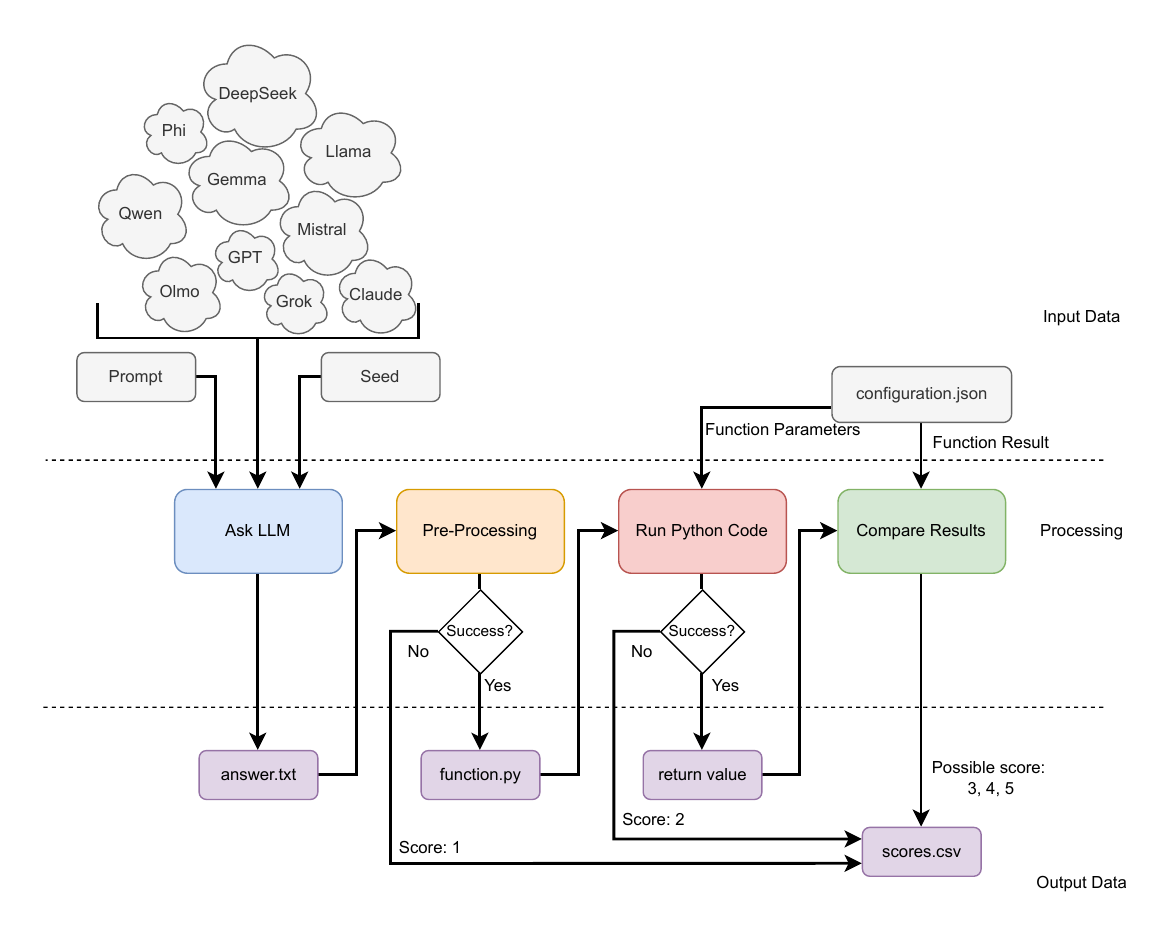}
    \caption{Evaluation pipeline for assessing LLM-generated Python code. Each prompt is submitted to an LLM with a random number generator seed, followed by code extraction, execution, and validation. Outputs are scored based on success at each stage, from parsing to result accuracy, enabling reproducible and structured performance assessment.}
    \label{fig:block_architecture}
\end{figure}

The process of evaluating LLM-generated Python code is depicted in Fig.~\ref{fig:block_architecture}. For each experiment, all combinations of predefined LLMs, random seeds, and prompts are systematically iterated. Each prompt is submitted to the respective LLM, and the resulting output is stored.

Function extraction from LLM responses proceeds in up to two steps: (1) code is extracted from the first Markdown code block, specifically from the text enclosed between the \texttt{\textasciigrave\textasciigrave\textasciigrave{}python} opening delimiter and the closing
\texttt{\textasciigrave\textasciigrave\textasciigrave{}} (that is, the code is expected to be located within these code fences); or (2) if the previous step fails, the entire response is assessed as potential Python code. If neither approach yields valid code, the attempt is logged and assigned a score of~1, as illustrated in Fig.~\ref{fig:block_architecture}. Both prompts explicitly instruct the LLMs to output a single code block containing only the function implementation; nonetheless, these extraction steps provide tolerance for minor deviations from reply formatting. Upon successful extraction, the function is saved as a Python file for subsequent execution.

Extracted code is then executed in a controlled environment. In Experiment~1 (\textit{ParShift}), the extracted \texttt{do\_parshift\_exp()} function is tested using three CSV files, collecting the respective turn-usurping proportions. In Experiment~2, \texttt{do\_clustering\_exp()} is executed with $n=30$ samples, with results collected as a \textit{Pandas} data frame. If execution fails due to syntax or runtime errors, the error in question is logged and a score of~2 is assigned.

Successful execution leads to an automated validation of the function's output type and format. For Experiment~1, the output must be a list of three floats; for Experiment~2, the expected output is a data frame with specific columns and dimensions. Mismatches in type or structure result in a score of~3. If the output format is correct, values are compared to pre-computed baselines: a 1\% numerical tolerance is allowed for Experiment~1, while statistical indistinguishability is required for Experiment~2 (see Section~\ref{sec:methods:datan} for details). Outputs deviating from the baseline receive a score of~4; otherwise, a perfect score of~5 is assigned.

This evaluation procedure is summarized as follows:

\begin{enumerate}
    \item No Python code could be extracted (score~1).
    \item Code fails to run due to syntax or runtime errors (score~2).
    \item Code runs but returns an output of incorrect type or structure (score~3).
    \item Code runs and returns the correct type/structure, but with incorrect results (numerically or statistically different from baseline; score~4).
    \item Code runs and returns the correct, baseline-matching result (score~5).
\end{enumerate}

All outcomes, scores, and error logs are recorded for further analysis. This systematic, automated pipeline enables robust and reproducible benchmarking of the LLM-generated code.

\subsection{LLMs Considered}
\label{sec:methods:llms}

The language models evaluated in this work, summarized in Table~\ref{table:llms}, were selected for their impact in Artificial Intelligence research, diverse methodological approaches, and demonstrated strengths in code generation, reasoning, and efficiency. Geographic diversity was also prioritized to ensure broad representation of current LLM development. Model sizes (parameters, tokens, etc.)  are indicated in millions (M), billions (B), or trillions (T), with uppercase letters used throughout for consistency.

\begin{table}[!tb]
    \caption{LLMs tested in this study. `Patch/Date` indicates the model's exact release patch, if available, otherwise displaying the patch release date. `Size` indicates the number of parameters in billions (B), when available. `Mode` shows whether the model was executed offline via Ollama in the author's institution infrastructure, or if it was experimented with online through the respective parent company's infrastructure. `Tag` corresponds to how each specific model is mentioned in the figures and tables of the results section.}
    \label{table:llms}
    \centering

    \begin{tabular}{llllrll}
        \toprule
        \textbf{Family} &
        \textbf{Model/Version} &
        \textbf{Ref.} &
        \textbf{Patch/Date} &
        \multicolumn{1}{l}{\textbf{Size}} &
        \textbf{Mode} &
        \textbf{Tag} \\

        \midrule

        Claude &
        Sonnet 3.7 &
        \cite{anthropic2025claude37} &
        20250219 &
        $\star$ &
        \llmonline &
        \texttt{claude-3.7-sonnet} \\

        \midrule

        DeepSeek &
        coder-v2 &
        \cite{zhu2024deepseek} &
        2024-09-06 &
        16B &
        \llmoffline &
        \texttt{deepseek-coder-v2}  \\

        &
        R1 &
        \cite{guo2025deepseekr1} &
        2025-01-20 &
        671B &
        \llmonline &
        \texttt{deepseek-r1} \\

        &
        V3 &
        \cite{liu2024deepseekv3} &
        0324 &
        671B &
        \llmonline &
        \texttt{deepseek-v3} \\

        \midrule

        Gemma &
        code &
        \cite{zhao2024codegemma} &
        2024-07-18 &
        7B &
        \llmoffline & \texttt{codegemma} \\

        &
        3.0 &
        \cite{kamath2025gemma3} &
        2025-04-18 &
        27B &
        \llmoffline &
        \texttt{gemma3} \\

        \midrule

        GPT &
        4o &
        \cite{hurst2024gpt} &
        2024-08-06 &
        $\star$ &
        \llmonline &
        \texttt{gpt-4o} \\

        &
        4.1 &
        \cite{openai2025gpt41} &
        2025-04-14 &
        $\star$ &
        \llmonline &
        \texttt{gpt-4.1} \\

        \midrule

        Grok &
        3-beta &
        \cite{xai2025grok3beta} &
        2025-02-17 &
        $\star$ &
        \llmonline &
        \texttt{grok-3-beta} \\

        \midrule

        LLaMA &
        3.3 &
        \cite{grattafiori2024llama} &
        2024-12-06 &
        70B &
        \llmoffline &
        \texttt{llama3.3}  \\

        &
        code &
        \cite{roziere2023code} &
        2024-07-18 &
        70B &
        \llmoffline &
        \texttt{codellama} \\

        \midrule

        Mistral &
        codestral 0.1 &
        \cite{mistral2024codestral} &
        2024-09-03 &
        22B &
        \llmoffline &
        \texttt{codestral} \\

        &
        Large 2.1 &
        \cite{mistral2024large} &
        24.11 &
        123B &
        \llmonline &
        \texttt{mistral-large} \\

        \midrule

        Olmo &
        2 &
        \cite{olmo20242} &
        2025-01-11&
        13B &
        \llmoffline &
        \texttt{olmo2} \\

        \midrule

        Phi &
        4.0 &
        \cite{abdin2024phi} &
        2025-01-08 &
        14B &
        \llmoffline &
        \texttt{phi4} \\

        \midrule

        Qwen &
        2.5-coder &
        \cite{hui2025qwen25coder} &
        2025-05-28 &
        32B &
        \llmoffline &
        \texttt{qwen2.5-coder} \\

        &
        3.0 &
        \cite{yang2025qwen3} &
        2025-05-29 &
        32B &
        \llmoffline &
        \texttt{qwen3} \\

        \bottomrule
        \multicolumn{6}{l}{\footnotesize{\textsuperscript{$\star$}Not publicly disclosed.}}

    \end{tabular}
\end{table}

Claude 3.7 Sonnet~\cite{anthropic2025claude37}, developed by Anthropic, is a state-of-the-art multimodal LLM capable of processing both textual and visual input. The model supports both rapid-response and extended-reasoning modes and is optimized for handling complex instruction-following, including coding tasks and natural language understanding. Claude 3.7 Sonnet is a representative advanced general-purpose LLM, demonstrating strong performance in both standard and zero-shot Python code synthesis tasks.

DeepSeek Coder‑V2~\cite{zhu2024deepseek}, R1~\cite{guo2025deepseekr1}, and V3~\cite{liu2024deepseekv3} belong to DeepSeek's open source LLM lineup. Coder‑V2 is an Mixture-of-Experts (MoE) model specialized for code and mathematics, supporting 338 programming languages. V3 is DeepSeek's 671B MoE general-purpose model, employing multi-head latent attention and reinforcement learning (RL) from human feedback for multilingual reasoning. R1 is a dense reasoning model derived from V3 using multi-stage RL---initially via pure RL and later with cold‑start supervised data---to improve math, logic, and code reasoning. The inclusion of these models demonstrates how conditional computation (via MoE) and scale benefit Python code generation, especially in zero‑shot scenarios with sparse online code examples.

CodeGemma~\cite{zhao2024codegemma} and Gemma3~\cite{kamath2025gemma3} are based on Google's lightweight Gemma architecture. CodeGemma (7B) is fine-tuned specifically for code completion and delivers strong performance on Python benchmarks, while Gemma3 (27B) is a general-purpose, multimodal model featuring long context windows and instruction tuning. These models allow examination of how compact, optimized open weight architectures perform in zero-shot code generation, especially when integrating lesser-known library APIs.

GPT‑4o~\cite{hurst2024gpt} and GPT‑4.1~\cite{openai2025gpt41} are successive generations of OpenAI's flagship LLMs. GPT‑4o (``omni'') is a unified multimodal model for text, image, audio, and video, trained end-to-end and achieving leading text and code performance with improved multilingual and latency characteristics. GPT‑4.1 is oriented towards developer applications, with a million-token context window and substantial improvements in code generation. Including both enables direct comparison between generalist multimodal (GPT-4o) and developer-optimized (GPT-4.1) architectures in zero-shot Python code synthesis with uncommon libraries.

Grok~3~\cite{xai2025grok3beta}, developed by xAI, is notable for its exceptionally large context window, officially claimed to support up to 1M tokens. This extended context capacity enables the model to capture long-range dependencies, which is valuable for complex code generation tasks. The model's advanced reasoning abilities and long-context handling make it a relevant benchmark for zero-shot application of novel libraries.

Llama 3.3~\cite{grattafiori2024llama} and CodeLlama~\cite{roziere2023code}, both from Meta, share a decoder-only transformer backbone but diverge in specialization. Llama 3.3 is a multilingual, instruction-tuned model with strong general-language and reasoning capabilities, while CodeLlama is fine-tuned for code---particularly Python---demonstrating strong performance at infilling and instruction-following. Their comparison illustrates the effect of generalist versus code-specialized tuning on Python code generation with rarely used APIs.

Codestral~\cite{mistral2024codestral} and Mistral Large 2.1~\cite{mistral2024large}, from Mistral AI, represent recent advances in both code-focused and general-purpose transformer models from Europe. Codestral is a compact (22B parameter) model designed specifically for code generation, supporting over 80 programming languages and offering competitive performance and latency on code benchmarks. Mistral Large 2.1 is a large (123B parameter), dense model with a 128K token context window, demonstrating strong reasoning and code generation capabilities. Including both allows assessment of architectural diversity and regional competitiveness in LLM-based code generation.

Olmo2~\cite{olmo20242} (13B), from the Allen Institute for AI, is a fully open source dense transformer trained on approximately 5T tokens. Its rigorous engineering and transparent training pipeline distinguish it among open models, supporting robust reasoning and general language understanding. Olmo2 offers a benchmark for transparent, research-driven LLMs in zero-shot Python code synthesis with less-common libraries.

Phi-4~\cite{abdin2024phi}, a 14B parameter model from Microsoft Research, exemplifies a data-centric training approach, leveraging high-quality synthetic data throughout pretraining and post‑training to advance reasoning, mathematical problem-solving, and code generation. Phi-4's strong performance relative to larger models---as also observed in a previous study from our group~\cite{fernandes2025deepseek}---underlines the influence of training methodology over sheer parameter count, and its compact size makes it well-suited for resource- or latency-constrained deployment.

Qwen2.5‑Coder~\cite{hui2025qwen25coder} and Qwen3~\cite{yang2025qwen3}, developed by Alibaba Cloud, expand the study's geographic and architectural breadth. Qwen2.5‑Coder, a dense model with a 128K-token context window, is reported to perform well in code generation tasks through synthetic and code-centric pretraining. Qwen3 offers both dense and MoE variants, dual-mode inference, adaptive compute, and multilingual reasoning, achieving competitive results in code and logic tasks. Together with the DeepSeek models, the inclusion of Alibaba's LLMs provides insight into the capabilities of open-weight LLMs from Asia in Python code generation when guided by prompts that include explicit API docstrings.

While every effort was made to include a broad range of state-of-the-art models, certain recent releases were excluded due to practical limitations in time, budget, hardware, or technical access. For example, Gemini models were omitted due to payment system issues with Google during the study period, and Llama4 exceeded available local hardware capacity. The rapid proliferation of LLMs precludes exhaustive coverage in a single study. Nonetheless, the selection here captures a representative cross-section of contemporary architectures, parameter scales, technical strategies, and geographic origins, providing a robust foundation for evaluating Python code generation---particularly in the context of novel Python libraries.

\subsection{Experimental Setup}
\label{sec:methods:expsetup}

To evaluate the capabilities of the models described in Subsection~\ref{sec:methods:llms}, the prompts detailed in Subsection~\ref{sec:methods:scenarios} were submitted to each LLM in six separate runs, using distinct pseudo-random number generator seeds where supported. While seed control is intended to improve reproducibility, not all LLMs provide this functionality (e.g., Claude 3.7 Sonnet did not support seeding at the time of writing), and even when seeding is available, strict reproducibility cannot be assured. Sources of non-determinism include parallel computation on heterogeneous hardware (such as multi-GPU configurations), small discrepancies in floating-point arithmetic across hardware types and architectures, and potential variability arising from system drivers or firmware. Further, instability in the software environment---including differences in library versions, system configuration, and computational resource allocation---can also impact reproducibility~\cite{semmelrock2024reproducibility,zhou2025assessingmacromicroeffects}. Accordingly, this study focuses on assessing the general statistical consistency of model outputs over six runs per configuration, rather than pursuing exact reproducibility throughout all models and conditions. As a consequence, random seeds are used on a best-effort basis, with the recognition that exact reproducibility may not be achievable under these conditions.

The temperature parameter for sampling was set to 10\% of each model's maximum supported temperature (e.g., $T=0.1$ if $T_\text{max}=1.0$), as this is generally considered adequate to allow for some stochasticity while still maintaining accurate problem-solving capabilities in code generation~\cite{renze2024effect,fernandes2025deepseek}. It should be noted, however, that not all models expose or respect temperature controls. For example, DeepSeek R1 (accessed via its online API) ignored temperature settings at the time of experimentation~\cite{dsr12025api}. Furthermore, the documentation for some models is ambiguous regarding the maximum temperature value. For instance, Mistral's official sampling guide illustrates temperatures up to 3.2~\cite{mistral2025sampling}, but in the present study a value of 0.2 was used, following Mistral's recommendations and to maintain consistency with other models employing a 0.0--2.0 range.

Other model parameters, such as \texttt{top\_p} (which restricts next-token sampling to the most probable candidates summing to a specified cumulative probability), were left at their default values.
Offline models were executed locally via Ollama~\cite{ollama2025} on the authors' institutional infrastructure, while 
online models were accessed through their official APIs. 

For models supporting external tools---such as sandboxed code execution or web search (e.g., GPT-4o and GPT-4.1)---these functionalities were not enabled or activated. All models were therefore evaluated solely on their intrinsic capabilities in processing the prompt and associated API docstrings, without access to additional external information sources beyond what was explicitly provided within each prompt.

\subsection{Data Collection and Analysis}
\label{sec:methods:datan}

For each experiment, the scores assigned to each LLM-generated function (ranging from 1 to 5) were analyzed to provide an overview of model performance over runs with different random seeds. Score distributions were visualized using histograms, allowing for a clear depiction of the relative frequency of each outcome per model over six runs. In addition to these distributions, the percentage of top scores (i.e., the proportion of runs yielding a perfect score of 5) was recorded, as this directly reflects the frequency with which each model succeeded in generating a fully correct solution.

To compare the proportion of correct answers between models, pairwise statistical testing was performed using Fisher's exact test~\cite{fisher1922interpretation}. For each model, its proportion of perfect scores (score equal to 5) was compared against that of each other model, and the number of statistically significant ``wins'' was recorded. Statistical significance is defined as a $p$-value below the conventional threshold of $\alpha = 0.05$. To account for multiple comparisons, $p$-values were adjusted using the Benjamini–Hochberg procedure for controlling the false discovery rate (FDR)~\cite{benjamini1995controlling}, applied per model (i.e., to each model's set of pairwise comparisons).

In Experiment~2, to further investigate the quality of runnable LLM-generated functions---specifically, those able to return properly formatted outputs---their results were statistically compared to those of the baseline \textit{pyclugen} experiment. Following the output comparison approach described in~\cite{fachada2017model}, the outputs of each function ($V$-measure as a function of average cluster length for each clustering algorithm, with centered and scaled results concatenated by algorithm) were subjected to principal component analysis (PCA). Statistical similarity was assessed using the Mann–Whitney $U$ test~\cite{mann1947test} on the first principal component scores. In addition, PCA score plots were visualized to provide a qualitative assessment of potential differences between LLM-generated and baseline outputs.

In addition to functional correctness, the subset of fully correct code snippets (score~5) was further analyzed for code quality using four software engineering metrics: cyclomatic complexity ($c_c$), maintainability index ($m_i$), number of type errors per 100 source lines of code ($e_t/100$), and number of \textit{F}~errors per 100 source lines of code ($e_F/100$). Source lines of code ($s_\mathrm{loc}$), defined as the number of non-comment, non-blank lines, were also calculated to characterize code size and to normalize error counts; these values were obtained using \textit{Radon}~\cite{radon2012}, a static analysis tool for Python. Cyclomatic complexity, also computed with \textit{Radon}, quantifies the number of independent paths through a program and provides a measure of structural complexity~\cite{mccabe1976complexity}, which is particularly relevant for assessing whether generated code remains simple and usable. The maintainability index, $m_i$, likewise measured with \textit{Radon}, is a composite indicator on a 0--100 scale (higher is better) derived from $c_c$, $s_\mathrm{loc}$, percentage of comment lines, and various measures of operators and operands. In this context it provides an aggregate view of readability and long-term maintainability. Type errors were identified using \textit{mypy}~\cite{mypy2014}, a static type checker for Python, with results expressed as $e_t/100$ to account for code size and allow comparison between models; these errors indicate violations of type annotations and potential runtime failures. \textit{F}~errors were determined using \textit{Ruff}~\cite{ruff2022}, a high-performance Python linter that extends and replaces \textit{Flake8}~\cite{flake82016}, detecting a broad range of issues including unused imports, undefined variables, duplicate definitions, and standard style violations. These issues are categorized under the traditional \textit{F} error codes, reflecting their role in capturing logical and correctness problems beyond type checking or complexity analysis. As with type errors, results were normalized as $e_F/100$ to mitigate differences in code size. For all metrics, emphasis was placed on median values, which are more robust to skewed distributions and outliers, while box plots were used to visualize overall distributions and highlight variability. This enabled a quantitative summary of correct LLM-generated code quality, allowing comparisons not only between models but also across experiments, providing insights into both model-specific tendencies and differences in code quality associated with task complexity.
Although there is some overlap between the metrics employed (e.g., cyclomatic complexity is one of the components of the maintainability index, and certain \textit{F}~errors may coincide with type-related issues), their joint use nevertheless provides a deeper assessment of code quality than any single metric alone.

Additional analyses are provided in the supplementary material~\cite{fachada2025suppl} for interested readers. These include further descriptive statistics (e.g., mean, standard deviation, among others), as well as pairwise model comparisons of full score distributions using the Mann–Whitney $U$ test (note that the main statistical analysis in the paper addresses the proportion of fully correct responses, score~$=5$, not the full score distribution). In addition, the supplementary material contains non-aggregated results and individual counts of type and \textit{F}~errors, allowing a more fine-grained inspection of model behavior. These supplementary analyses are not further discussed in the main text since: 1) descriptive statistics such as means or standard deviations may obscure important aspects of discrete, skewed, or multi-modal distributions; 2) while overall scores are useful to understand how and when models fail, emphasizing statistically significant differences in the complete score distribution between models may distract from the primary criterion of interest: the consistent generation of fully correct code; and, 3) detailed per-instance code quality results, although useful for in-depth inspection, extend beyond the central focus of this work on functional correctness.

Data analysis was conducted primarily in Python, using several established scientific computing libraries. \textit{Pandas}~\cite{mckinney2011pandas} and \textit{NumPy}~\cite{harris2020array} were employed for general data manipulation and numerical computations, while \textit{SciPy}~\cite{virtanen2020scipy} provided statistical testing functionalities. The \textit{statsmodels} package~\cite{seabold2010statsmodels} was used to perform $p$-value corrections. 
Additionally, some analyses were performed in the R environment~\cite{r2025r}. In particular, the \textit{micompr} package~\cite{fachada2016micompr} was used for statistically comparing multidimensional outputs from the LLM-generated functions in Experiment~2 against a baseline, enabling differentiation between functions scoring 4 and those scoring 5. The specific versions of these packages are listed in Table~\ref{tab:softvers}.

\section{Results}
\label{sec:results}

Table~\ref{tab:scores} summarizes the aggregated evaluation results for the 17 LLMs tested in the two experiments. Detailed pairwise model comparisons are provided in Table~\ref{tab:exp1_pvals} for Experiment~1 and Table~\ref{tab:exp2_pvals} for Experiment~2. Additionally, to distinguish between scores of 4 and 5 for Experiment~2, Table~\ref{tab:exp2_micomp} provides the statistical comparison of outputs from LLM-generated functions against a predefined baseline.

\newcommand{\histw}{0.9cm}

\begin{table}[!tb]
    \caption{Distribution and significance of scores (1--5) for the 17 tested models in the two experiments. `Hist.' shows the histogram (score distribution); `Top' reports the percentage of top scores (i.e., scores of 5); and `Sig.' indicates the number of models for which this model had a statistically significant higher proportion of top scores (Fisher's exact test~\cite{fisher1922interpretation}, $\alpha < 0.05$, with $p$-values corrected for FDR using the Benjamini--Hochberg method~\cite{benjamini1995controlling}; additional test details in Table~\ref{tab:exp1_pvals} for Experiment~1, and Table~\ref{tab:exp2_pvals} for Experiment~2).} 
    \label{tab:scores}
    \begin{tabular}{lcrrcrr}
        \toprule
        & \multicolumn{3}{l}{\textbf{Experiment~1}} & \multicolumn{3}{l}{\textbf{Experiment~2}} \\
        \cmidrule(l){2-4}
        \cmidrule(l){5-7}
        \textbf{Model} & \textbf{Hist.} & \multicolumn{1}{l}{\textbf{Top}} & \multicolumn{1}{l}{\textbf{Sig.}} & \textbf{Hist.} & \multicolumn{1}{l}{\textbf{Top}} & \multicolumn{1}{l}{\textbf{Sig.}}   \\
        \midrule

\texttt{claude-3.7-sonnet} & \resizebox{\histw}{!}{\drawHistogram{0}{4}{0}{0}{2}} &  33.3\% & 0 & \resizebox{\histw}{!}{\drawHistogram{0}{2}{0}{0}{4}} &  66.7\% & 10\\
\texttt{codegemma} & \resizebox{\histw}{!}{\drawHistogram{0}{6}{0}{0}{0}} &   0.0\% & 0 & \resizebox{\histw}{!}{\drawHistogram{0}{6}{0}{0}{0}} &   0.0\% & 0\\
\texttt{codellama} & \resizebox{\histw}{!}{\drawHistogram{0}{4}{2}{0}{0}} &   0.0\% & 0 & \resizebox{\histw}{!}{\drawHistogram{0}{6}{0}{0}{0}} &   0.0\% & 0\\
\texttt{codestral} & \resizebox{\histw}{!}{\drawHistogram{0}{2}{0}{0}{4}} &  66.7\% & 0 & \resizebox{\histw}{!}{\drawHistogram{0}{6}{0}{0}{0}} &   0.0\% & 0\\
\texttt{deepseek-coder-v2} & \resizebox{\histw}{!}{\drawHistogram{0}{6}{0}{0}{0}} &   0.0\% & 0 & \resizebox{\histw}{!}{\drawHistogram{0}{6}{0}{0}{0}} &   0.0\% & 0\\
\texttt{deepseek-r1} & \resizebox{\histw}{!}{\drawHistogram{0}{4}{0}{0}{2}} &  33.3\% & 0 & \resizebox{\histw}{!}{\drawHistogram{0}{2}{0}{0}{4}} &  66.7\% & 10\\
\texttt{deepseek-v3} & \resizebox{\histw}{!}{\drawHistogram{0}{2}{0}{0}{4}} &  66.7\% & 0 & \resizebox{\histw}{!}{\drawHistogram{0}{0}{4}{0}{2}} &  33.3\% & 0\\
\texttt{gemma3} & \resizebox{\histw}{!}{\drawHistogram{0}{0}{0}{6}{0}} &   0.0\% & 0 & \resizebox{\histw}{!}{\drawHistogram{0}{6}{0}{0}{0}} &   0.0\% & 0\\
\texttt{gpt-4.1} & \resizebox{\histw}{!}{\drawHistogram{0}{0}{0}{0}{6}} & 100.0\% & 12 & \resizebox{\histw}{!}{\drawHistogram{0}{0}{0}{0}{6}} & 100.0\% & 11\\
\texttt{gpt-4o} & \resizebox{\histw}{!}{\drawHistogram{0}{2}{0}{0}{4}} &  66.7\% & 0 & \resizebox{\histw}{!}{\drawHistogram{0}{1}{2}{0}{3}} &  50.0\% & 0\\
\texttt{grok-3-beta} & \resizebox{\histw}{!}{\drawHistogram{0}{0}{0}{0}{6}} & 100.0\% & 12 & \resizebox{\histw}{!}{\drawHistogram{0}{1}{0}{0}{5}} &  83.3\% & 10\\
\texttt{llama3.3} & \resizebox{\histw}{!}{\drawHistogram{0}{6}{0}{0}{0}} &   0.0\% & 0 & \resizebox{\histw}{!}{\drawHistogram{0}{6}{0}{0}{0}} &   0.0\% & 0\\
\texttt{mistral-large} & \resizebox{\histw}{!}{\drawHistogram{0}{4}{0}{0}{2}} &  33.3\% & 0 & \resizebox{\histw}{!}{\drawHistogram{0}{0}{0}{0}{6}} & 100.0\% & 11\\
\texttt{olmo2} & \resizebox{\histw}{!}{\drawHistogram{0}{6}{0}{0}{0}} &   0.0\% & 0 & \resizebox{\histw}{!}{\drawHistogram{0}{6}{0}{0}{0}} &   0.0\% & 0\\
\texttt{phi4} & \resizebox{\histw}{!}{\drawHistogram{0}{4}{0}{2}{0}} &   0.0\% & 0 & \resizebox{\histw}{!}{\drawHistogram{0}{1}{0}{5}{0}} &   0.0\% & 0\\
\texttt{qwen2.5-coder} & \resizebox{\histw}{!}{\drawHistogram{0}{6}{0}{0}{0}} &   0.0\% & 0 & \resizebox{\histw}{!}{\drawHistogram{0}{6}{0}{0}{0}} &   0.0\% & 0\\
\texttt{qwen3} & \resizebox{\histw}{!}{\drawHistogram{1}{3}{0}{2}{0}} &   0.0\% & 0 & \resizebox{\histw}{!}{\drawHistogram{2}{4}{0}{0}{0}} &   0.0\% & 0\\

        \bottomrule
    \end{tabular}
\end{table}

To clarify the nature of execution failures that resulted in a score of 2, the corresponding error messages were grouped into three categories (Table~\ref{tab:errors}). \textbf{Logic} errors refer to basic Python or reasoning flaws, such as missing imports or referencing variables before assignment. \textbf{Established API} errors arise from the incorrect use of widely adopted libraries such as \textit{Pandas} or \textit{scikit-learn}. Finally, \textbf{novel API} errors correspond to issues in handling the libraries under evaluation, specifically \textit{ParShift} in Experiment~1 and \textit{pyclugen} in Experiment~2. This categorization condenses a wide range of raw error messages into a small set of recurring patterns, highlighting the main reasons for LLM-generated code failure in these experiments.

\begin{table}[!tb]
\centering
\caption{Source of errors when code fails to run due to syntax or runtime errors (score 2) for the 17 tested models in the two experiments. `Logic' errors include basic Python or reasoning errors such as missing imports or use of variable before assignment; `Est. API' errors are due to incorrect use of established APIs such as \textit{Pandas} or \textit{scikit-learn}; finally, `Nov. API' errors occur when misusing the novel API under test, i.e., \textit{ParShift} in Experiment~1 and \textit{pyclugen} in Experiment~2.}
\label{tab:errors}
\begin{tabular}{lrrrrrr}
\toprule
 & \multicolumn{3}{l}{\textbf{Experiment 1}} & \multicolumn{3}{l}{\textbf{Experiment 2}} \\
\cmidrule(l){2-4}
\cmidrule(l){5-7}
\textbf{Model} & \multicolumn{1}{l}{\textbf{Logic}} & \multicolumn{1}{l}{\textbf{Est. API}} & \multicolumn{1}{l}{\textbf{Nov. API}} & \multicolumn{1}{l}{\textbf{Logic}} & \multicolumn{1}{l}{\textbf{Est. API}} & \multicolumn{1}{l}{\textbf{Nov. API}} \\
\midrule

\texttt{claude-3.7-sonnet} &	--	&	--	&	4	&	2	&	--	&	--	\\
\texttt{codegemma} &	--	&	--	&	6	&	--	&	6	&	--	\\
\texttt{codellama} &	--	&	--	&	4	&	--	&	1	&	5	\\
\texttt{codestral} &	--	&	--	&	2	&	--	&	6	&	--	\\
\texttt{deepseek-coder-v2} &	--	&	--	&	6	&	--	&	6	&	--	\\
\texttt{deepseek-r1} &	--	&	--	&	4	&	2	&	--	&	--	\\
\texttt{deepseek-v3} &	1	&	--	&	1	&	--	&	--	&	--	\\
\texttt{gemma3} &	--	&	--	&	--	&	--	&	6	&	--	\\
\texttt{gpt-4.1} &	--	&	--	&	--	&	--	&	--	&	--	\\
\texttt{gpt-4o} &	--	&	--	&	2	&	--	&	1	&	--	\\
\texttt{grok-3-beta} &	--	&	--	&	--	&	--	&	1	&	--	\\
\texttt{llama3.3} &	--	&	--	&	6	&	--	&	6	&	--	\\
\texttt{mistral-large} &	--	&	--	&	4	&	--	&	--	&	--	\\
\texttt{olmo2} &	6	&	--	&	--	&	--	&	6	&	--	\\
\texttt{phi4} &	--	&	--	&	4	&	--	&	1	&	--	\\
\texttt{qwen2.5-coder} &	--	&	--	&	6	&	--	&	6	&	--	\\
\texttt{qwen3} &	--	&	--	&	3	&	2	&	--	&	2	\\

\bottomrule
\end{tabular}
\end{table}

\begin{figure}[!tp]
\includegraphics[width=1\textwidth]{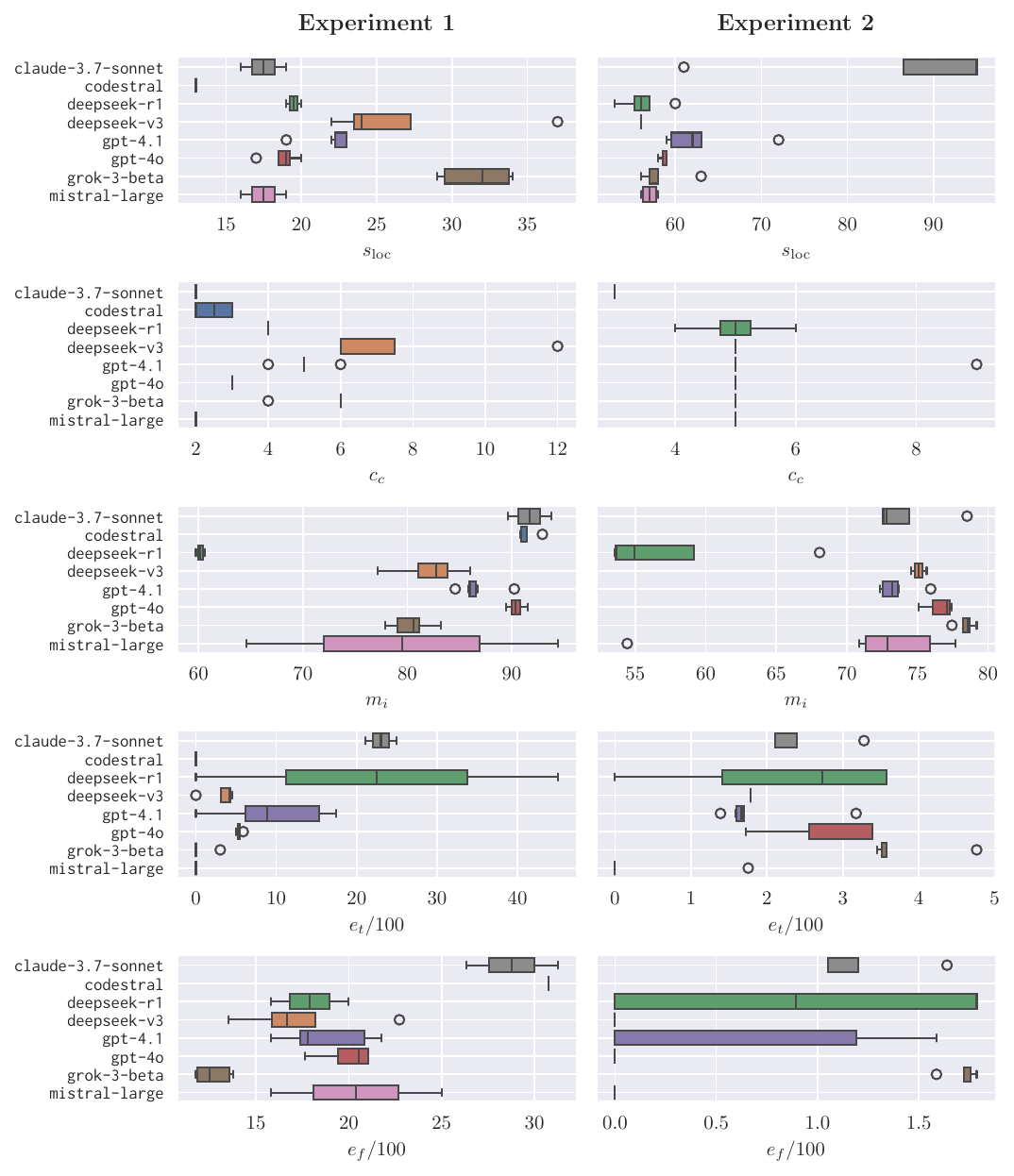}
\centering
\caption{Standard box plots of code quality metrics per model per experiment, restricted to code achieving score = 5. Left column: Experiment~1; right column: Experiment~2 (note that axis scales differ substantially between experiments and should not be compared directly). No box plot is shown for Codestral in Experiment~2, since this model did not generate working code. Rows correspond to different metrics: $s_\mathrm{loc}$---source lines of code, i.e., non-comment lines of code; $c_c$---cyclomatic complexity, a measure of independent code paths (lower is better); $m_i$---maintainability index, a composite software quality measure (0--100, higher is better); $e_t/100$---type errors per 100~$s_\mathrm{loc}$, reflecting type inconsistencies; and, $e_F/100$---\textit{F}~errors per 100~$s_\mathrm{loc}$, reflecting potential code quality issues.}
\label{fig:quality}
\end{figure}

To complement the functional correctness analysis, code quality indicators for the subset of runs that produced fully correct code (score~5) are also reported. Table~\ref{tab:quality} summarizes, for each model and experiment, the medians of $s_\mathrm{loc}$, $c_c$ (cyclomatic complexity), $m_i$ (maintainability index), $e_t/100$ (type errors per 100~$s_\mathrm{loc}$), and $e_F/100$ (\textit{F}~errors per 100~$s_\mathrm{loc}$), with $n$ indicating the number of successful instances. The last row (``All instances'') reports experiment-wide medians (and total $n$). Figure~\ref{fig:quality} complements this table by displaying standard box plots of these metrics per model and experiment, providing a clearer view of distributional spread and potential outliers.

\begin{table}[!tb]

\centering
\caption{Median code quality metrics per model for each experiment for runs that produced fully correct code (score 5): $s_\mathrm{loc}$---source lines of code, i.e., non-comment lines of code; $c_c$---cyclomatic complexity, a measure of independent code paths (lower is better); $m_i$---maintainability index, a composite software quality measure (0--100, higher is better); $e_t/100$---type errors per 100~$s_\mathrm{loc}$, reflecting type inconsistencies; and, $e_F/100$---\textit{F}~errors per 100~$s_\mathrm{loc}$, reflecting potential code quality issues. `All instances' shows experiment-wide medians of each metric across all models, with $n$ equal to the total number of successful instances.}
\label{tab:quality}
\setlength{\tabcolsep}{3.55pt}
\begin{tabular}{lrrrrrrrrrrrr}
\toprule
 & \multicolumn{6}{l}{\textbf{Experiment 1}} & \multicolumn{6}{l}{\textbf{Experiment 2}} \\
\cmidrule(l){2-7}
\cmidrule(l){8-13}
\textbf{Model} & $n$ & $s_\mathrm{loc}$ & $c_c$ & $m_i$ & $e_t/100$ & $e_F/100$ & $n$ & $s_\mathrm{loc}$ & $c_c$ & $m_i$ & $e_t/100$ & $e_F/100$ \\
\midrule

{\small\texttt{claude-3.7-son.}} & 2 & 17.5 & 2.0 & 91.7 & 23.03 & 28.78 & 4 & 95.0 & 3.0 & 72.8 & 2.11 & 1.05\\
{\small\texttt{codestral}} & 4 & 13.0 & 2.5 & 90.9 & 0.00 & 30.77 & -- & -- & -- & -- & -- & --\\
{\small\texttt{deepseek-r1}} & 2 & 19.5 & 4.0 & 60.2 & 22.50 & 17.89 & 4 & 56.0 & 5.0 & 54.9 & 2.73 & 0.89\\
{\small\texttt{deepseek-v3}} & 4 & 24.0 & 6.0 & 82.7 & 4.17 & 16.67 & 2 & 56.0 & 5.0 & 75.1 & 1.79 & 0.00\\
{\small\texttt{gpt-4.1}} & 6 & 23.0 & 5.0 & 86.0 & 8.89 & 17.79 & 6 & 62.0 & 5.0 & 73.2 & 1.67 & 0.00\\
{\small\texttt{gpt-4o}} & 4 & 19.0 & 3.0 & 90.4 & 5.26 & 20.53 & 3 & 59.0 & 5.0 & 77.1 & 3.39 & 0.00\\
{\small\texttt{grok-3-beta}} & 6 & 32.0 & 6.0 & 80.6 & 0.00 & 12.51 & 5 & 57.0 & 5.0 & 78.6 & 3.51 & 1.75\\
{\small\texttt{mistral-large}} & 2 & 17.5 & 2.0 & 79.5 & 0.00 & 20.39 & 6 & 57.0 & 5.0 & 72.9 & 0.00 & 0.00\\
\midrule
\textit{All instances} & 30 & 21.0 & 4.0 & 86.0 & 3.60 & 17.91 & 30 & 58.0 & 5.0 & 73.5 & 1.84 & 0.00\\

\bottomrule
\end{tabular}
\end{table}

Complete experimental data, including all generated outputs, extracted functions, and error logs, are available in the supplementary material~\cite{fachada2025suppl}.

\subsection{Experiment~1: \textit{ParShift}}

The first task involved generating Python code using the \textit{ParShift} library to compute the proportion of utterance pairs classified as ``Turn Usurping'' from CSV conversation files. As indicated in Table~\ref{tab:scores}, the models exhibited a wide range of outcomes. GPT-4.1 and Grok~3 received a perfect score of~5 in every evaluated run, generating executable Python code that returned the expected output in all cases. These two models also achieved a statistically significant higher proportion of perfect scores than other models in 12 out of 16 pairwise comparisons (see Table~\ref{tab:exp1_pvals}).

Other high-performing models include Codestral, DeepSeek-V3, and GPT-4o, which generated correct solutions in four out of six runs. Claude 3.7 Sonnet, DeepSeek-R1, and Mistral Large achieved score~5 results in two out of six runs each. All large models executed online achieved at least one perfect result. Among offline models, only Codestral reached a perfect score, doing so in 66.7\% of runs. Other offline models did not produce any perfect scores in this experiment.

Interestingly, Gemma3 always produced code that executed without errors, but returned an incorrect---although properly formatted---result in every run (score~4). 
As summarized in Table~\ref{tab:errors}, most execution failures (score~2) were novel API errors. These typically manifested as \texttt{KeyError}s when attempting to access non-existent \textit{ParShift} fields, observed in models such as Llama~3.3, Phi-4, Qwen2.5-Coder, DeepSeek-R1, Mistral Large, and Codestral.  In several cases, exceptions surfaced from \textit{Pandas} within \textit{ParShift}'s I/O pipeline (mainly in DeepSeek Coder-V2), but these were triggered by incorrect use of \textit{ParShift}'s API and were therefore classified as novel API errors. A secondary source of failures involved logic errors, including missing or incorrect imports, as was always the case in Olmo2. There were no direct established API errors in this experiment.

As can be observed in Table~\ref{tab:quality} and Fig.~\ref{fig:quality}, the characteristics of the LLM-generated code that achieved score~5 differed substantially across models in Experiment~1. Regarding size, Codestral-generated code was concise and uniform (median $s_\mathrm{loc}=13$, without variance), while code from Grok 3 was comparatively longer (typically $>30$ lines). Structural complexity was low for code produced by Claude 3.7 Sonnet and Mistral Large (median $c_c=2$), while code from Grok 3 and DeepSeek-V3 showed higher branching (median $c_c>5$), including an outlier for DeepSeek-V3 with $c_c=12$. Maintainability was high for code generated by Claude 3.7 Sonnet, Codestral, and GPT-4o (median $m_i>90$), whereas code from DeepSeek-R1 showed lower maintainability (median $m_i\approx60$). Type checking outcomes varied: several models yielded zero or near-zero $e_t/100$, but code from Claude 3.7 Sonnet and DeepSeek-R1 exhibited higher medians ($>20$ type errors per 100~$s_\mathrm{loc}$). \textit{F}~errors were mostly related with unused inputs and a few inconsequential type redefinitions. Code from Grok~3 had the lowest median $(\approx12.5$ per 100~$s_\mathrm{loc}$), while code from Claude 3.7 Sonnet and Codestral showed higher medians (roughly 28--31 per 100~$s_\mathrm{loc}$). Focusing on models with 100\% success rate, GPT-4.1 stands out by always being on the average of other models in all these metrics, and Grok~3 by producing somewhat verbose and complex code, although with very few type and \textit{F}~errors.

\subsection{Experiment~2: \textit{pyclugen}}

The second experiment required the generation of a function to perform a clustering experiment using the \textit{pyclugen} and \textit{scikit-learn} libraries and return a \textit{Pandas} data frame with computed clustering metrics. We hypothesize that this task is more complex than that of Experiment~1 due to the higher dimensionality of the data and the need to properly apply multiple clustering algorithms.

The GPT-4.1 and Mistral Large models produced correct, executable code that returned the expected output in every run, obtaining a perfect score of~5 in all cases. The Grok 3 model produced correct output in five out of six runs, while Claude 3.7 Sonnet and DeepSeek-R1 succeeded in four out of six runs. These five models showed a statistically significant higher proportion of top scores relative to most others (details in Table~\ref{tab:exp2_pvals}).

The remaining online models, GPT-4o and DeepSeek-V3, produced correct solutions in some runs, but none of the smaller, offline models succeeded in generating correct code in any evaluated run. This observation is in line with our hypothesis that Experiment~2 involves greater complexity than Experiment~1.

The Phi-4 model consistently produced code that executed successfully and returned a valid data frame in the correct format, but the results were statistically different from those provided by the reference implementation in all tested runs (see Table~\ref{tab:exp2_micomp}). This was due to the generated code applying clustering algorithms to the point projections (which \textit{pyclugen} uses as an intermediate step to generate the final points) instead of the final data points themselves, resulting in structurally correct but semantically incorrect outputs.

As shown in Table~\ref{tab:errors}, the most common code execution failures in Experiment~2 (score~2) were established API errors, mainly from incorrect use of \textit{scikit-learn}. A recurring issue in the code generated by several models (CodeGemma, Codestral, DeepSeek Coder-V2, Gemma3, Llama3.3, Olmo2, and Qwen2.5-Coder) was importing \texttt{GaussianMixture} from \texttt{sklearn.cluster} instead of the correct \texttt{sklearn.mixture} module. Additional problems included passing deprecated/invalid parameters in \textit{scikit-learn} methods (CodeLlama) and accessing unavailable attributes in \textit{scikit-learn} objects (Grok~3), as well as \textit{Pandas} usage incompatible with recent versions (GPT-4o and Phi-4). Novel API errors were less frequent but present, reflecting misuse of \textit{pyclugen}, such as calling the \texttt{clugen()} function with missing required positional arguments (CodeLlama), or assuming non-existent functions or attributes (CodeLlama and Qwen3). Finally, a smaller set of logic errors was observed, including missing imports or symbols (Claude 3.7 Sonnet and DeepSeek-R1) and failure to generate the experiment's function with the specified name (Qwen3).

In addition to these execution failures, some models produced outputs with the wrong format (score~3), in particular returning a data frame with the column name ``ahc-average'' rather than the required ``ahc-avg'' (DeepSeek-V3 and GPT-4o).

Focusing only on fully functional generated code (score~5), broadly similar profiles were observed between models, with a few interesting differences, as shown in Table~\ref{tab:quality} and Fig.~\ref{fig:quality}. In terms of size, code from Claude~3.7~Sonnet was markedly more verbose (median $s_\mathrm{loc}=95$), whereas other models clustered around 56--62 lines. Cyclomatic complexity was stable (median $c_c=5$) for all models except Claude~3.7~Sonnet (median $c_c=2$); the combination of higher $s_\mathrm{loc}$ and lower $c_c$ suggests longer but more linear implementations with fewer branching constructs. An isolated outlier was observed for GPT-4.1 with $c_c=7$. Maintainability indices were generally similar between models (medians in the 70s), with lower values for code from DeepSeek-R1 (median $m_i\approx55$) and an occasional low outlier for Mistral~Large (below~55). Type checking results were uniformly strong: medians of $e_t/100$ were low for all models, including a zero median for Mistral~Large; code from Grok~3 showed the highest median among models ($\approx3.5$ per 100~$s_\mathrm{loc}$), though still small in absolute terms. \textit{F}~errors per 100~$s_\mathrm{loc}$ ($e_F/100$) were also minimal, and consistently zero for DeepSeek-V3, GPT-4o, and Mistral~Large. The few observed \textit{F}~errors were related to undefined types (with no impact on runtime execution) and some assigned variables that were never used. Code from Grok~3 showed the highest median ($\approx1.75$ per 100~$s_\mathrm{loc}$). Among models achieving perfect correctness across all runs, code from GPT-4.1 tended again to align with group medians across metrics, whereas code from Mistral~Large frequently reached zero $e_t/100$ and $e_F/100$.

\subsection{Aggregate Observations}

In both experiments, GPT-4.1 was the only model to achieve a perfect score in every evaluated run. Grok 3 also performed strongly, with only one instance of a non-executable output in Experiment~2. At the other end of the scale, Qwen3 was the only model to produce score~1 outputs, failing to generate a Python function at all in some cases. Concurrently, score~4 was commonly assigned to Gemma3 (in Experiment~1) and Phi-4 (in Experiment~2) due to correct function execution but semantically incorrect results.

In both experiments, most outputs clustered around either score~2 (runtime or syntax error) or score~5 (correct result). Scores of~3 (e.g., incorrect return type or number of elements) and~4 (incorrect results with valid output) were less frequent. Within the score~2 outputs, Experiment~1 failures were largely due to novel API errors (mainly from incorrectly accessing \textit{ParShift}'s data structures), whereas Experiment~2 was dominated by established API errors (misuse or hallucination of methods, mainly in \textit{scikit-learn}).

Considering the last row of Table~\ref{tab:quality}, both experiments yielded the same number of successful instances ($n=30$). Code size increased substantially from Experiment~1 to Experiment~2 (median $s_\mathrm{loc}$ from $21$ to $58$). Median cyclomatic complexity rose only modestly (median $c_c$ from $4$ to $5$) and was notably more stable in Experiment~2, with values tightly concentrated around $5$. The maintainability index decreased between experiments (median $m_i$ from $86$ to $73.5$), consistent with the greater breadth of the second task. Error rates per 100~$s_\mathrm{loc}$ were lower in Experiment~2: the median $e_t/100$ was roughly halved ($1.84$ vs.\ $3.60$), and the median $e_F/100$ dropped from approximately $18$ to $0$.

\section{Discussion}
\label{sec:discussion}

The results obtained from the two experiments provide insight into the capabilities and limitations of LLMs in generating Python code using third-party libraries, particularly when dealing with less commonly known APIs. Several important observations can be drawn from the analysis of results.

In Experiment~1, most LLMs encountered difficulties with the \textit{ParShift} library, resulting predominantly from key errors when accessing its internal data structures. These failures can be traced in part to limitations in the clarity of the existing \textit{ParShift} documentation. Indeed, during the preliminary phase of this study, the authors themselves faced challenges in clearly formulating the final prompt due to ambiguities and complexities in the library's documentation. A post-experiment assessment of \textit{ParShift}'s documentation confirms that clearer structure and more explicit examples would likely reduce such errors. Consequently, improvements to \textit{ParShift}'s documentation are planned, motivated directly by findings from this study.

In Experiment~2, the incorrect import of \texttt{GaussianMixture} was a common failure point for many models, reflecting a broader class of import and attribute errors. This issue highlights a fundamental characteristic of LLMs: their reliance on statistical patterns in training data to predict plausible tokens rather than verifying correctness through explicit validation. Since \texttt{GaussianMixture} appears contextually similar to algorithms within the \texttt{sklearn.cluster} module, models often incorrectly associated it with this module. This behavior resembles human guesswork based on contextual association rather than verified knowledge, reflecting a known limitation of current generative models~\cite{bender2020climbing}. In contrast, models such as GPT-4.1 and Mistral Large consistently avoided this mistake, suggesting that more rigorous fine-tuning with validated code samples, stronger internal consistency mechanisms, or filtering of incorrect token sequences for critical libraries such as \textit{scikit-learn} may mitigate such errors.

Code quality results demonstrated that top performers are not identical once maintainability and static issues are considered. GPT-4.1 tended to sit near group medians across metrics---a ``steady default'' that consistently produced working code, albeit not always optimizing for maintainability or strict style conformance. Its counterpart, GPT-4o, was also relatively average across metrics but showed higher maintainability when it produced correct outputs. Grok~3 was notably verbose in Experiment~1 yet exhibited very few type and \textit{F}~errors; conversely, in Experiment~2 it was among the less verbose models while showing comparatively higher static error rates among successful runs. A plausible explanation is that the simpler task in Experiment~1 involved a more idiosyncratic API (\textit{ParShift}) that several models found harder to parse, whereas the more complex Experiment~2 may have benefited from clearer API affordances (e.g., \textit{pyclugen} documentation), allowing some models to adapt structure to the increased task complexity, with only a modest reduction in maintainability. Importantly, Grok~3 almost always produced working code (score~5) across both tasks, trailing only GPT-4.1 in overall reliability. Claude~3.7~Sonnet presented the lowest and most stable $c_c$ between experiments. It achieved this with relatively few $s_\textrm{loc}$ in Experiment~1 but with substantially higher $s_\textrm{loc}$ in Experiment~2 (approximately 95 versus 56--62 for peers), yielding longer, more linear code that may be easier to read, while maintaining reasonable $m_i$ and a relatively higher static error rate in Experiment~1. Codestral, although only successful in Experiment~1, produced concise code with low $c_c$, high $m_i$, and zero type errors. However, it exhibited the highest median $e_F/100$, largely attributable to unused (often type-related) imports---suggesting a conservative import strategy or remnants of partially pruned templates. DeepSeek-R1 showed consistently lower $m_i$ relative to peers in both experiments, indicating a higher refactoring burden even when outputs were correct. Its non-reasoning counterpart DeepSeek-V3 generated code with higher maintainability and fewer static issues, indicating that R1 did not surpass V3 in this study. Finally, while Mistral Large struggled in Experiment~1 (both in success rate and some quality indicators), it excelled in Experiment~2, producing correct code in all runs with uniformly strong quality metrics (e.g., zero type and \textit{F} errors).

With the exception of Codestral, models that produced working code in one experiment also did so in the other. Across successful runs, Experiment~2 generally yielded longer code (higher $s_\textrm{loc}$), only slightly higher $c_c$, and a modest decrease in $m_i$---patterns consistent with increased task complexity. Interestingly, type and \textit{F} errors were markedly fewer in Experiment~2 among correct solutions, which, together with the error taxonomy, strengthens the hypothesis that \textit{ParShift}'s API in Experiment~1 posed interpretability challenges for several models. This observation complements the earlier conclusion regarding the value of improved documentation for \textit{ParShift}.

During prompt development for Experiment~2, a separate issue arose when some LLMs provided \texttt{np.int64} types for the \texttt{seed} parameter in the \texttt{clugen()} function from the \textit{pyclugen} library. Although conceptually correct, this caused runtime crashes due to type incompatibility with the previous implementation of \texttt{pyclugen}. To address this, \texttt{pyclugen} was updated prior to the experiment to accept any integer-compatible type, ensuring robustness to typical coding practices. This experience highlights a broader utility of employing LLMs for testing assumptions and identifying edge cases that might not surface even under comprehensive unit testing. In this particular instance, despite 100\% unit test coverage, the practical use of LLM-generated code facilitated discovery and rectification of a subtle but impactful type-checking oversight.

Contrasting with the findings from a previous study~\cite{fernandes2025deepseek}, where smaller offline models, particularly Phi-4 and Llama3.3, performed comparably to larger models, the current results revealed significantly lower performance from these smaller models in generating code for computational experiments. The exception among offline models was Codestral, which achieved a few instances of correct results in Experiment~1. The diminished performance of Phi-4 and Llama3.3 compared to previous findings indicates that the complexity and novelty of the APIs used in the present study significantly affect the effectiveness of these models.

Among all evaluated models, GPT-4.1 demonstrated a clear and consistent advantage, achieving perfect scores in both experiments and showing a strong capacity to interpret unfamiliar APIs and return syntactically and semantically correct code. This suggests that recent improvements in the GPT-4 family, such as improved reasoning, instruction following, and error correction mechanisms, translate directly into more reliable code synthesis, especially for complex or lesser-known libraries. Both Grok 3 and Mistral Large also performed at a high level, ranking among the few models able to solve both tasks with high consistency. These results indicate that, at present, only a handful of top-tier models can reliably perform advanced code generation tasks that involve integrating novel APIs in end-to-end experiment pipelines.

It is also relevant that DeepSeek-R1, a model explicitly marketed for improved reasoning, did not clearly outperform its generalist counterpart, DeepSeek-V3, in these scenarios, especially in Experiment~1. Both models achieved similar levels of performance and operated in a comparable range to Claude~3.7~Sonnet and GPT-4o, which are also positioned as high-capacity but general-purpose LLMs. This outcome suggests that, despite recent advances in LLM architectures and training strategies for specialized reasoning, practical gains in challenging code synthesis tasks may remain limited unless accompanied by targeted fine-tuning and extensive exposure to domain-specific code samples. The high variance observed even among top models further underlines the need for ongoing evaluation and benchmarking as LLM technology continues to evolve.

For settings where multiple models achieve similar top-score rates, code quality metrics can guide selection according to context: models with lower $e_t/100$ and $e_F/100$ may be preferable when static typing and automated code checks (continuous integration linting gates that block errors before code is merged) are critical. Higher $s_\textrm{loc}$ with lower $c_c$ may be acceptable (or even desirable) when readability by less experienced developers is prioritized; and ``median-leaning'' profiles can be advantageous when consistency across tasks and codebases is valued.

Overall, these findings highlight the importance of rigorous API documentation and model-specific fine-tuning when employing LLMs for automated generation of computational experiment code. Additionally, these results demonstrate the utility of such automated code generation tasks not only for evaluating model capabilities but also for exposing assumptions and deficiencies within libraries themselves, thus contributing positively to their further development and robustness. The integration of code quality metrics highlights that, beyond attaining functional correctness, the sustainability and maintainability of LLM-generated code should be factored into model selection and deployment decisions.

\section{Limitations}
\label{sec:limitations}

While the present study provides important insights into the capabilities of current LLMs for the automated synthesis of computational experiment code using less common Python libraries, several limitations should be acknowledged.

First, the evaluation was limited to two specific computational experiments involving the \textit{ParShift} and \textit{pyclugen} libraries. While these libraries are representative of real-world, lesser-known scientific software, the generalizability of findings to other domains or libraries cannot be guaranteed. It is possible that the observed failure patterns or success rates may differ when different APIs, programming languages, or experimental paradigms are used. Additionally, because the prompts included real docstrings from the target libraries, we cannot fully exclude the possibility that these materials were part of some LLMs' pretraining corpora, potentially influencing their performance.

Second, all models were evaluated under a zero-shot prompt setting, which, while arguably reflecting realistic scenarios faced by many users, does not capture the potential benefits of in-context learning, prompt tuning, or multi-turn interactions. Many recent studies show that LLMs can substantially improve code quality when provided with examples, clarifying questions, or iterative feedback~\cite{chen2021evaluating,10.1145/3510003.3510203,gu2024effectiveness}. The single-pass prompt structure employed here, while controlled, does not fully exploit these capabilities.

Third, the evaluation infrastructure did not simulate the availability of external resources (e.g., internet search, code validation tools, or documentation lookup) that are increasingly integrated into commercial LLMs and coding assistants. All models were evaluated using only their pretrained knowledge. Therefore, the results may underestimate the real-world performance of models when used in modern coding environments.

Fourth, model variability was assessed by running each prompt with several seeds and temperature slightly above zero to guarantee some randomness; however, this does not exhaustively capture the full space of model stochasticity, especially in models for which temperature or seed control was not available or did not function as expected. Thus, some sources of non-determinism and rare failure modes may not have been observed.

Finally, the study included a limited set of models, selected for breadth and relevance but necessarily omitting some recent or proprietary releases due to practical access constraints. The results may not reflect the capabilities of models released after the completion of this study, nor those of specialized, commercial, or domain-adapted variants.

These limitations point to important avenues for future research, such as expanding the diversity of tested libraries and tasks, incorporating iterative and in-context prompting, 
and benchmarking LLMs in environments augmented with external tools and resources.

\section{Conclusions}
\label{sec:conclusions}

This study systematically evaluated the ability of state-of-the-art LLMs to generate functional Python code for realistic computational experiments involving less widely known scientific libraries. By presenting models with structured, zero-shot prompts using the \textit{ParShift} and \textit{pyclugen} libraries, the experiments probed both code synthesis accuracy and robustness when integrating novel APIs.

The results reveal that only a small subset of current models, particularly GPT-4.1, Grok~3, and Mistral Large, consistently produced correct and executable code across both experiments. These models demonstrated strong capabilities not only in generating syntactically correct code, but also in interpreting prompt requirements and accurately handling library-specific details. Among runs that achieved perfect functional scores, analysis of code quality indicators (source lines of code, cyclomatic complexity, maintainability index, and static error rates) further differentiated models and exposed practical trade-offs in readability and maintainability, highlighting the value of assessing quality beyond pass/fail correctness. Interestingly, the specialized reasoning model DeepSeek-R1 did not clearly surpass the generalist DeepSeek-V3, and both performed similarly to Claude~3.7~Sonnet and GPT-4o. Conversely, the majority of smaller or open weight offline models struggled, often failing at basic integration steps or producing code with semantic errors.

Beyond model assessment, this study also highlighted the role of LLM-based experiments in uncovering both documentation gaps (as seen with \textit{ParShift}) and latent software bugs (as demonstrated by the type handling issue in \textit{pyclugen}), suggesting the mutual benefit of LLM evaluation for both model developers and scientific software authors.

While the results are encouraging regarding the best-performing LLMs, the observed limitations and failure cases indicate that reliably automating scientific experiment pipelines with LLMs remains challenging---particularly for models lacking targeted fine-tuning or advanced context handling. To broaden the applicability of LLMs in specialized research domains, continued progress is needed not only in model training and prompt engineering, but also in improving the documentation of third-party libraries. Clearer and more complete API documentation can help LLMs interpret unfamiliar tools more accurately, reducing errors and improving overall performance.

Future work should extend these benchmarks to a wider range of libraries, languages, and interaction modes, as well as to real-world use cases requiring code maintenance, interpretability, and safe deployment. As LLMs continue to advance, systematic and transparent evaluation will remain critical for their responsible and effective integration into scientific research workflows.

\section*{Data Availability}

The data generated by this study and its respective analysis are available at \url{https://doi.org/10.5281/zenodo.16582736} under the CC-BY license.

\section*{Acknowledgements}

This research was partially funded by the Funda\c c\~ao para a Ci\^encia e a Tecnologia (FCT, \url{https://ror.org/00snfqn58}) under Grants UIDB/04111/2020, UIDB/00066/2020, \linebreak
UIDB/00408/2020, CEECINST/00002/2021/CP2788/CT0001 and the LASIGE Research Unit, ref. UID/000408/2025, as well as by the Instituto Lusófono de Investigação e Desenvolvimento (ILIND), Portugal, under Project COFAC/ILIND\-/\-COPELABS\-/1/2024.

\appendix
\section{Supplementary Tables}
\label{app1}

\renewcommand{\thetable}{A.\arabic{table}}
\setcounter{table}{0}

This appendix contains four supplementary tables. Table~\ref{tab:softvers} provides the versions of the software used for running the LLM-generated code and performing the data analysis. The remaining tables provide detailed statistical analyses of the experimental results. Specifically, Table~\ref{tab:exp1_pvals} presents pairwise $p$-values for top-score proportions in Experiment~1, Table~\ref{tab:exp2_pvals} presents the corresponding results for Experiment~2, and Table~\ref{tab:exp2_micomp} provides a comparison of model outputs for Experiment~2 against the reference baseline.

\begin{table}[!htpb]
\caption{Software versions used for executing LLM-generated code (column `Execution') and performing the data analysis, including static code quality assessment (column `Analysis').}\label{tab:softvers}
\centering
\begin{tabular}{lcc}
\toprule
\textbf{Software} & \textbf{Execution} & \textbf{Analysis} \\
\midrule
Python & 3.10.12 & 3.12.3 \\
\textit{mypy} & --- & 1.17.1 \\
\textit{NumPy} & 1.26.4 & 2.3.2 \\
\textit{Pandas} & 2.2.3 & 2.3.1 \\
\textit{ParShift} & 1.0.1 & --- \\
\textit{pyclugen} & 1.1.4 & --- \\
\textit{Radon} & --- & 6.0.1 \\
\textit{Ruff} & --- & 0.12.9 \\
\textit{scikit-learn} & 1.4.2 & --- \\
\textit{SciPy} & --- & 1.16.1 \\
\textit{statsmodels} & --- & 0.14.5 \\
\midrule
R & --- & 4.5.1 \\
\textit{micompr} & --- & 1.2.0\\
\textit{rmarkdown} & --- & 2.29\\
\bottomrule
\end{tabular}
\end{table}

\begin{sidewaystable}
    \caption{Experiment~1: $p$-values from Fisher's exact test~\cite{fisher1922interpretation} comparing the proportion of top scores between models (rows vs. columns). The null hypothesis $H_0$ states that the model in the row does \textit{not} have a significantly higher proportion of top scores than the model in the column. $p$-values below the 0.05 significance threshold (highlighted in light grey) indicate rejection of $H_0$, suggesting that the row model \textit{does} have a significantly higher proportion of top scores. Multiple comparisons are corrected per row using the Benjamini--Hochberg procedure~\cite{benjamini1995controlling} over 16 comparisons.} 
    \label{tab:exp1_pvals}
    {
    \setlength{\tabcolsep}{4pt}
    \begin{tabular}{rrrrrrrrrrrrrrrrrr} 
        \toprule
        {\small \textbf{Model}} & 

        \rotatebox{90}{\texttt{\footnotesize claude-3.7-sonnet}} & 
        \rotatebox{90}{\texttt{\footnotesize codegemma}} & 
        \rotatebox{90}{\texttt{\footnotesize codellama}} & 
        \rotatebox{90}{\texttt{\footnotesize codestral}} & 
        \rotatebox{90}{\texttt{\footnotesize deepseek-coder-v2}} & 
        \rotatebox{90}{\texttt{\footnotesize deepseek-r1}} & 
        \rotatebox{90}{\texttt{\footnotesize deepseek-v3}} & 
        \rotatebox{90}{\texttt{\footnotesize gemma3}} & 
        \rotatebox{90}{\texttt{\footnotesize gpt-4.1}} & 
        \rotatebox{90}{\texttt{\footnotesize gpt-4o}} & 
        \rotatebox{90}{\texttt{\footnotesize grok-3-beta}} & 
        \rotatebox{90}{\texttt{\footnotesize llama3.3}} & 
        \rotatebox{90}{\texttt{\footnotesize mistral-large}} & 
        \rotatebox{90}{\texttt{\footnotesize olmo2}} & 
        \rotatebox{90}{\texttt{\footnotesize phi4}} & 
        \rotatebox{90}{\texttt{\footnotesize qwen2.5-coder}} & 
        \rotatebox{90}{\texttt{\footnotesize qwen3}} \\       
        
        \midrule

\texttt{\footnotesize claude-3.7-sonnet} & --- & {\footnotesize 0.404} & {\footnotesize 0.404} & {\footnotesize 1.000} & {\footnotesize 0.404} & {\footnotesize 1.000} & {\footnotesize 1.000} & {\footnotesize 0.404} & {\footnotesize 1.000} & {\footnotesize 1.000} & {\footnotesize 1.000} & {\footnotesize 0.404} & {\footnotesize 1.000} & {\footnotesize 0.404} & {\footnotesize 0.404} & {\footnotesize 0.404} & {\footnotesize 0.404}\\
\texttt{\footnotesize codegemma} & {\footnotesize 1.000} & --- & {\footnotesize 1.000} & {\footnotesize 1.000} & {\footnotesize 1.000} & {\footnotesize 1.000} & {\footnotesize 1.000} & {\footnotesize 1.000} & {\footnotesize 1.000} & {\footnotesize 1.000} & {\footnotesize 1.000} & {\footnotesize 1.000} & {\footnotesize 1.000} & {\footnotesize 1.000} & {\footnotesize 1.000} & {\footnotesize 1.000} & {\footnotesize 1.000}\\
\texttt{\footnotesize codellama} & {\footnotesize 1.000} & {\footnotesize 1.000} & --- & {\footnotesize 1.000} & {\footnotesize 1.000} & {\footnotesize 1.000} & {\footnotesize 1.000} & {\footnotesize 1.000} & {\footnotesize 1.000} & {\footnotesize 1.000} & {\footnotesize 1.000} & {\footnotesize 1.000} & {\footnotesize 1.000} & {\footnotesize 1.000} & {\footnotesize 1.000} & {\footnotesize 1.000} & {\footnotesize 1.000}\\
\texttt{\footnotesize codestral} & {\footnotesize 0.378} & {\footnotesize 0.054} & {\footnotesize 0.054} & --- & {\footnotesize 0.054} & {\footnotesize 0.378} & {\footnotesize 0.831} & {\footnotesize 0.054} & {\footnotesize 1.000} & {\footnotesize 0.831} & {\footnotesize 1.000} & {\footnotesize 0.054} & {\footnotesize 0.378} & {\footnotesize 0.054} & {\footnotesize 0.054} & {\footnotesize 0.054} & {\footnotesize 0.054}\\
\texttt{\footnotesize deepseek-coder-v2} & {\footnotesize 1.000} & {\footnotesize 1.000} & {\footnotesize 1.000} & {\footnotesize 1.000} & --- & {\footnotesize 1.000} & {\footnotesize 1.000} & {\footnotesize 1.000} & {\footnotesize 1.000} & {\footnotesize 1.000} & {\footnotesize 1.000} & {\footnotesize 1.000} & {\footnotesize 1.000} & {\footnotesize 1.000} & {\footnotesize 1.000} & {\footnotesize 1.000} & {\footnotesize 1.000}\\
\texttt{\footnotesize deepseek-r1} & {\footnotesize 1.000} & {\footnotesize 0.404} & {\footnotesize 0.404} & {\footnotesize 1.000} & {\footnotesize 0.404} & --- & {\footnotesize 1.000} & {\footnotesize 0.404} & {\footnotesize 1.000} & {\footnotesize 1.000} & {\footnotesize 1.000} & {\footnotesize 0.404} & {\footnotesize 1.000} & {\footnotesize 0.404} & {\footnotesize 0.404} & {\footnotesize 0.404} & {\footnotesize 0.404}\\
\texttt{\footnotesize deepseek-v3} & {\footnotesize 0.378} & {\footnotesize 0.054} & {\footnotesize 0.054} & {\footnotesize 0.831} & {\footnotesize 0.054} & {\footnotesize 0.378} & --- & {\footnotesize 0.054} & {\footnotesize 1.000} & {\footnotesize 0.831} & {\footnotesize 1.000} & {\footnotesize 0.054} & {\footnotesize 0.378} & {\footnotesize 0.054} & {\footnotesize 0.054} & {\footnotesize 0.054} & {\footnotesize 0.054}\\
\texttt{\footnotesize gemma3} & {\footnotesize 1.000} & {\footnotesize 1.000} & {\footnotesize 1.000} & {\footnotesize 1.000} & {\footnotesize 1.000} & {\footnotesize 1.000} & {\footnotesize 1.000} & --- & {\footnotesize 1.000} & {\footnotesize 1.000} & {\footnotesize 1.000} & {\footnotesize 1.000} & {\footnotesize 1.000} & {\footnotesize 1.000} & {\footnotesize 1.000} & {\footnotesize 1.000} & {\footnotesize 1.000}\\
\texttt{\footnotesize gpt-4.1} & {\footnotesize \cellcolor{black!10}0.040} & {\footnotesize \cellcolor{black!10}$<$0.01} & {\footnotesize \cellcolor{black!10}$<$0.01} & {\footnotesize 0.242} & {\footnotesize \cellcolor{black!10}$<$0.01} & {\footnotesize \cellcolor{black!10}0.040} & {\footnotesize 0.242} & {\footnotesize \cellcolor{black!10}$<$0.01} & --- & {\footnotesize 0.242} & {\footnotesize 1.000} & {\footnotesize \cellcolor{black!10}$<$0.01} & {\footnotesize \cellcolor{black!10}0.040} & {\footnotesize \cellcolor{black!10}$<$0.01} & {\footnotesize \cellcolor{black!10}$<$0.01} & {\footnotesize \cellcolor{black!10}$<$0.01} & {\footnotesize \cellcolor{black!10}$<$0.01}\\
\texttt{\footnotesize gpt-4o} & {\footnotesize 0.378} & {\footnotesize 0.054} & {\footnotesize 0.054} & {\footnotesize 0.831} & {\footnotesize 0.054} & {\footnotesize 0.378} & {\footnotesize 0.831} & {\footnotesize 0.054} & {\footnotesize 1.000} & --- & {\footnotesize 1.000} & {\footnotesize 0.054} & {\footnotesize 0.378} & {\footnotesize 0.054} & {\footnotesize 0.054} & {\footnotesize 0.054} & {\footnotesize 0.054}\\
\texttt{\footnotesize grok-3-beta} & {\footnotesize \cellcolor{black!10}0.040} & {\footnotesize \cellcolor{black!10}$<$0.01} & {\footnotesize \cellcolor{black!10}$<$0.01} & {\footnotesize 0.242} & {\footnotesize \cellcolor{black!10}$<$0.01} & {\footnotesize \cellcolor{black!10}0.040} & {\footnotesize 0.242} & {\footnotesize \cellcolor{black!10}$<$0.01} & {\footnotesize 1.000} & {\footnotesize 0.242} & --- & {\footnotesize \cellcolor{black!10}$<$0.01} & {\footnotesize \cellcolor{black!10}0.040} & {\footnotesize \cellcolor{black!10}$<$0.01} & {\footnotesize \cellcolor{black!10}$<$0.01} & {\footnotesize \cellcolor{black!10}$<$0.01} & {\footnotesize \cellcolor{black!10}$<$0.01}\\
\texttt{\footnotesize llama3.3} & {\footnotesize 1.000} & {\footnotesize 1.000} & {\footnotesize 1.000} & {\footnotesize 1.000} & {\footnotesize 1.000} & {\footnotesize 1.000} & {\footnotesize 1.000} & {\footnotesize 1.000} & {\footnotesize 1.000} & {\footnotesize 1.000} & {\footnotesize 1.000} & --- & {\footnotesize 1.000} & {\footnotesize 1.000} & {\footnotesize 1.000} & {\footnotesize 1.000} & {\footnotesize 1.000}\\
\texttt{\footnotesize mistral-large} & {\footnotesize 1.000} & {\footnotesize 0.404} & {\footnotesize 0.404} & {\footnotesize 1.000} & {\footnotesize 0.404} & {\footnotesize 1.000} & {\footnotesize 1.000} & {\footnotesize 0.404} & {\footnotesize 1.000} & {\footnotesize 1.000} & {\footnotesize 1.000} & {\footnotesize 0.404} & --- & {\footnotesize 0.404} & {\footnotesize 0.404} & {\footnotesize 0.404} & {\footnotesize 0.404}\\
\texttt{\footnotesize olmo2} & {\footnotesize 1.000} & {\footnotesize 1.000} & {\footnotesize 1.000} & {\footnotesize 1.000} & {\footnotesize 1.000} & {\footnotesize 1.000} & {\footnotesize 1.000} & {\footnotesize 1.000} & {\footnotesize 1.000} & {\footnotesize 1.000} & {\footnotesize 1.000} & {\footnotesize 1.000} & {\footnotesize 1.000} & --- & {\footnotesize 1.000} & {\footnotesize 1.000} & {\footnotesize 1.000}\\
\texttt{\footnotesize phi4} & {\footnotesize 1.000} & {\footnotesize 1.000} & {\footnotesize 1.000} & {\footnotesize 1.000} & {\footnotesize 1.000} & {\footnotesize 1.000} & {\footnotesize 1.000} & {\footnotesize 1.000} & {\footnotesize 1.000} & {\footnotesize 1.000} & {\footnotesize 1.000} & {\footnotesize 1.000} & {\footnotesize 1.000} & {\footnotesize 1.000} & --- & {\footnotesize 1.000} & {\footnotesize 1.000}\\
\texttt{\footnotesize qwen2.5-coder} & {\footnotesize 1.000} & {\footnotesize 1.000} & {\footnotesize 1.000} & {\footnotesize 1.000} & {\footnotesize 1.000} & {\footnotesize 1.000} & {\footnotesize 1.000} & {\footnotesize 1.000} & {\footnotesize 1.000} & {\footnotesize 1.000} & {\footnotesize 1.000} & {\footnotesize 1.000} & {\footnotesize 1.000} & {\footnotesize 1.000} & {\footnotesize 1.000} & --- & {\footnotesize 1.000}\\
\texttt{\footnotesize qwen3} & {\footnotesize 1.000} & {\footnotesize 1.000} & {\footnotesize 1.000} & {\footnotesize 1.000} & {\footnotesize 1.000} & {\footnotesize 1.000} & {\footnotesize 1.000} & {\footnotesize 1.000} & {\footnotesize 1.000} & {\footnotesize 1.000} & {\footnotesize 1.000} & {\footnotesize 1.000} & {\footnotesize 1.000} & {\footnotesize 1.000} & {\footnotesize 1.000} & {\footnotesize 1.000} & ---\\

        \bottomrule
    \end{tabular}
    }
\end{sidewaystable}

\begin{sidewaystable}
    \caption{Experiment~2: $p$-values from Fisher's exact test~\cite{fisher1922interpretation} comparing the proportion of top scores between models (rows vs. columns). The null hypothesis $H_0$ states that the model in the row does \textit{not} have a significantly higher proportion of top scores than the model in the column. $p$-values below the 0.05 significance threshold (highlighted in light grey) indicate rejection of $H_0$, suggesting that the row model \textit{does} have a significantly higher proportion of top scores. Multiple comparisons are corrected per row using the Benjamini--Hochberg procedure~\cite{benjamini1995controlling} over 16 comparisons.} 
    \label{tab:exp2_pvals}
    {
    \setlength{\tabcolsep}{4pt}
    \begin{tabular}{rrrrrrrrrrrrrrrrrr} 
        \toprule
        {\small \textbf{Model}}  & 

        \rotatebox{90}{\texttt{\footnotesize claude-3.7-sonnet}} & 
        \rotatebox{90}{\texttt{\footnotesize codegemma}} & 
        \rotatebox{90}{\texttt{\footnotesize codellama}} & 
        \rotatebox{90}{\texttt{\footnotesize codestral}} & 
        \rotatebox{90}{\texttt{\footnotesize deepseek-coder-v2}} & 
        \rotatebox{90}{\texttt{\footnotesize deepseek-r1}} & 
        \rotatebox{90}{\texttt{\footnotesize deepseek-v3}} & 
        \rotatebox{90}{\texttt{\footnotesize gemma3}} & 
        \rotatebox{90}{\texttt{\footnotesize gpt-4.1}} & 
        \rotatebox{90}{\texttt{\footnotesize gpt-4o}} & 
        \rotatebox{90}{\texttt{\footnotesize grok-3-beta}} & 
        \rotatebox{90}{\texttt{\footnotesize llama3.3}} & 
        \rotatebox{90}{\texttt{\footnotesize mistral-large}} & 
        \rotatebox{90}{\texttt{\footnotesize olmo2}} & 
        \rotatebox{90}{\texttt{\footnotesize phi4}} & 
        \rotatebox{90}{\texttt{\footnotesize qwen2.5-coder}} & 
        \rotatebox{90}{\texttt{\footnotesize qwen3}} \\       
        
        \midrule

\texttt{\footnotesize claude-3.7-sonnet} & --- & {\footnotesize \cellcolor{black!10}0.048} & {\footnotesize \cellcolor{black!10}0.048} & {\footnotesize \cellcolor{black!10}0.048} & {\footnotesize \cellcolor{black!10}0.048} & {\footnotesize 0.895} & {\footnotesize 0.412} & {\footnotesize \cellcolor{black!10}0.048} & {\footnotesize 1.000} & {\footnotesize 0.667} & {\footnotesize 1.000} & {\footnotesize \cellcolor{black!10}0.048} & {\footnotesize 1.000} & {\footnotesize \cellcolor{black!10}0.048} & {\footnotesize \cellcolor{black!10}0.048} & {\footnotesize \cellcolor{black!10}0.048} & {\footnotesize \cellcolor{black!10}0.048}\\
\texttt{\footnotesize codegemma} & {\footnotesize 1.000} & --- & {\footnotesize 1.000} & {\footnotesize 1.000} & {\footnotesize 1.000} & {\footnotesize 1.000} & {\footnotesize 1.000} & {\footnotesize 1.000} & {\footnotesize 1.000} & {\footnotesize 1.000} & {\footnotesize 1.000} & {\footnotesize 1.000} & {\footnotesize 1.000} & {\footnotesize 1.000} & {\footnotesize 1.000} & {\footnotesize 1.000} & {\footnotesize 1.000}\\
\texttt{\footnotesize codellama} & {\footnotesize 1.000} & {\footnotesize 1.000} & --- & {\footnotesize 1.000} & {\footnotesize 1.000} & {\footnotesize 1.000} & {\footnotesize 1.000} & {\footnotesize 1.000} & {\footnotesize 1.000} & {\footnotesize 1.000} & {\footnotesize 1.000} & {\footnotesize 1.000} & {\footnotesize 1.000} & {\footnotesize 1.000} & {\footnotesize 1.000} & {\footnotesize 1.000} & {\footnotesize 1.000}\\
\texttt{\footnotesize codestral} & {\footnotesize 1.000} & {\footnotesize 1.000} & {\footnotesize 1.000} & --- & {\footnotesize 1.000} & {\footnotesize 1.000} & {\footnotesize 1.000} & {\footnotesize 1.000} & {\footnotesize 1.000} & {\footnotesize 1.000} & {\footnotesize 1.000} & {\footnotesize 1.000} & {\footnotesize 1.000} & {\footnotesize 1.000} & {\footnotesize 1.000} & {\footnotesize 1.000} & {\footnotesize 1.000}\\
\texttt{\footnotesize deepseek-coder-v2} & {\footnotesize 1.000} & {\footnotesize 1.000} & {\footnotesize 1.000} & {\footnotesize 1.000} & --- & {\footnotesize 1.000} & {\footnotesize 1.000} & {\footnotesize 1.000} & {\footnotesize 1.000} & {\footnotesize 1.000} & {\footnotesize 1.000} & {\footnotesize 1.000} & {\footnotesize 1.000} & {\footnotesize 1.000} & {\footnotesize 1.000} & {\footnotesize 1.000} & {\footnotesize 1.000}\\
\texttt{\footnotesize deepseek-r1} & {\footnotesize 0.895} & {\footnotesize \cellcolor{black!10}0.048} & {\footnotesize \cellcolor{black!10}0.048} & {\footnotesize \cellcolor{black!10}0.048} & {\footnotesize \cellcolor{black!10}0.048} & --- & {\footnotesize 0.412} & {\footnotesize \cellcolor{black!10}0.048} & {\footnotesize 1.000} & {\footnotesize 0.667} & {\footnotesize 1.000} & {\footnotesize \cellcolor{black!10}0.048} & {\footnotesize 1.000} & {\footnotesize \cellcolor{black!10}0.048} & {\footnotesize \cellcolor{black!10}0.048} & {\footnotesize \cellcolor{black!10}0.048} & {\footnotesize \cellcolor{black!10}0.048}\\
\texttt{\footnotesize deepseek-v3} & {\footnotesize 1.000} & {\footnotesize 0.364} & {\footnotesize 0.364} & {\footnotesize 0.364} & {\footnotesize 0.364} & {\footnotesize 1.000} & --- & {\footnotesize 0.364} & {\footnotesize 1.000} & {\footnotesize 1.000} & {\footnotesize 1.000} & {\footnotesize 0.364} & {\footnotesize 1.000} & {\footnotesize 0.364} & {\footnotesize 0.364} & {\footnotesize 0.364} & {\footnotesize 0.364}\\
\texttt{\footnotesize gemma3} & {\footnotesize 1.000} & {\footnotesize 1.000} & {\footnotesize 1.000} & {\footnotesize 1.000} & {\footnotesize 1.000} & {\footnotesize 1.000} & {\footnotesize 1.000} & --- & {\footnotesize 1.000} & {\footnotesize 1.000} & {\footnotesize 1.000} & {\footnotesize 1.000} & {\footnotesize 1.000} & {\footnotesize 1.000} & {\footnotesize 1.000} & {\footnotesize 1.000} & {\footnotesize 1.000}\\
\texttt{\footnotesize gpt-4.1} & {\footnotesize 0.260} & {\footnotesize \cellcolor{black!10}$<$0.01} & {\footnotesize \cellcolor{black!10}$<$0.01} & {\footnotesize \cellcolor{black!10}$<$0.01} & {\footnotesize \cellcolor{black!10}$<$0.01} & {\footnotesize 0.260} & {\footnotesize \cellcolor{black!10}0.044} & {\footnotesize \cellcolor{black!10}$<$0.01} & --- & {\footnotesize 0.121} & {\footnotesize 0.533} & {\footnotesize \cellcolor{black!10}$<$0.01} & {\footnotesize 1.000} & {\footnotesize \cellcolor{black!10}$<$0.01} & {\footnotesize \cellcolor{black!10}$<$0.01} & {\footnotesize \cellcolor{black!10}$<$0.01} & {\footnotesize \cellcolor{black!10}$<$0.01}\\
\texttt{\footnotesize gpt-4o} & {\footnotesize 1.000} & {\footnotesize 0.145} & {\footnotesize 0.145} & {\footnotesize 0.145} & {\footnotesize 0.145} & {\footnotesize 1.000} & {\footnotesize 0.727} & {\footnotesize 0.145} & {\footnotesize 1.000} & --- & {\footnotesize 1.000} & {\footnotesize 0.145} & {\footnotesize 1.000} & {\footnotesize 0.145} & {\footnotesize 0.145} & {\footnotesize 0.145} & {\footnotesize 0.145}\\
\texttt{\footnotesize grok-3-beta} & {\footnotesize 0.571} & {\footnotesize \cellcolor{black!10}0.012} & {\footnotesize \cellcolor{black!10}0.012} & {\footnotesize \cellcolor{black!10}0.012} & {\footnotesize \cellcolor{black!10}0.012} & {\footnotesize 0.571} & {\footnotesize 0.176} & {\footnotesize \cellcolor{black!10}0.012} & {\footnotesize 1.000} & {\footnotesize 0.364} & --- & {\footnotesize \cellcolor{black!10}0.012} & {\footnotesize 1.000} & {\footnotesize \cellcolor{black!10}0.012} & {\footnotesize \cellcolor{black!10}0.012} & {\footnotesize \cellcolor{black!10}0.012} & {\footnotesize \cellcolor{black!10}0.012}\\
\texttt{\footnotesize llama3.3} & {\footnotesize 1.000} & {\footnotesize 1.000} & {\footnotesize 1.000} & {\footnotesize 1.000} & {\footnotesize 1.000} & {\footnotesize 1.000} & {\footnotesize 1.000} & {\footnotesize 1.000} & {\footnotesize 1.000} & {\footnotesize 1.000} & {\footnotesize 1.000} & --- & {\footnotesize 1.000} & {\footnotesize 1.000} & {\footnotesize 1.000} & {\footnotesize 1.000} & {\footnotesize 1.000}\\
\texttt{\footnotesize mistral-large} & {\footnotesize 0.260} & {\footnotesize \cellcolor{black!10}$<$0.01} & {\footnotesize \cellcolor{black!10}$<$0.01} & {\footnotesize \cellcolor{black!10}$<$0.01} & {\footnotesize \cellcolor{black!10}$<$0.01} & {\footnotesize 0.260} & {\footnotesize \cellcolor{black!10}0.044} & {\footnotesize \cellcolor{black!10}$<$0.01} & {\footnotesize 1.000} & {\footnotesize 0.121} & {\footnotesize 0.533} & {\footnotesize \cellcolor{black!10}$<$0.01} & --- & {\footnotesize \cellcolor{black!10}$<$0.01} & {\footnotesize \cellcolor{black!10}$<$0.01} & {\footnotesize \cellcolor{black!10}$<$0.01} & {\footnotesize \cellcolor{black!10}$<$0.01}\\
\texttt{\footnotesize olmo2} & {\footnotesize 1.000} & {\footnotesize 1.000} & {\footnotesize 1.000} & {\footnotesize 1.000} & {\footnotesize 1.000} & {\footnotesize 1.000} & {\footnotesize 1.000} & {\footnotesize 1.000} & {\footnotesize 1.000} & {\footnotesize 1.000} & {\footnotesize 1.000} & {\footnotesize 1.000} & {\footnotesize 1.000} & --- & {\footnotesize 1.000} & {\footnotesize 1.000} & {\footnotesize 1.000}\\
\texttt{\footnotesize phi4} & {\footnotesize 1.000} & {\footnotesize 1.000} & {\footnotesize 1.000} & {\footnotesize 1.000} & {\footnotesize 1.000} & {\footnotesize 1.000} & {\footnotesize 1.000} & {\footnotesize 1.000} & {\footnotesize 1.000} & {\footnotesize 1.000} & {\footnotesize 1.000} & {\footnotesize 1.000} & {\footnotesize 1.000} & {\footnotesize 1.000} & --- & {\footnotesize 1.000} & {\footnotesize 1.000}\\
\texttt{\footnotesize qwen2.5-coder} & {\footnotesize 1.000} & {\footnotesize 1.000} & {\footnotesize 1.000} & {\footnotesize 1.000} & {\footnotesize 1.000} & {\footnotesize 1.000} & {\footnotesize 1.000} & {\footnotesize 1.000} & {\footnotesize 1.000} & {\footnotesize 1.000} & {\footnotesize 1.000} & {\footnotesize 1.000} & {\footnotesize 1.000} & {\footnotesize 1.000} & {\footnotesize 1.000} & --- & {\footnotesize 1.000}\\
\texttt{\footnotesize qwen3} & {\footnotesize 1.000} & {\footnotesize 1.000} & {\footnotesize 1.000} & {\footnotesize 1.000} & {\footnotesize 1.000} & {\footnotesize 1.000} & {\footnotesize 1.000} & {\footnotesize 1.000} & {\footnotesize 1.000} & {\footnotesize 1.000} & {\footnotesize 1.000} & {\footnotesize 1.000} & {\footnotesize 1.000} & {\footnotesize 1.000} & {\footnotesize 1.000} & {\footnotesize 1.000} & ---\\
        
        \bottomrule
    \end{tabular}
    }
\end{sidewaystable}

\begin{sidewaystable}
    \caption{Statistical comparisons of Experiment~2 baseline versus outputs from LLM-generated functions. The table shows the adjusted $p$-values (`$p$-val.') resulting from the Mann-Whitney U test~\cite{mann1947test} applied to the first principal component, obtained via PCA, of the concatenated algorithm outputs ($V$-measure per average cluster length, centered and scaled). The $p$-values are adjusted for FDR per-model (across seeds) using the Benjamini-Hochberg procedure~\cite{benjamini1995controlling}. Significant results (at $\alpha=0.05$) indicating statistical differences between baseline and LLM-generated outputs are highlighted with a light grey background. A significant difference implies assigning a score of 4 (statistically different), whereas non-significant results indicate assigning a score of 5 (statistically indistinguishable). The `sp' columns contain corresponding stylized score plots displaying the first two principal components to assist in evaluating statistical similarities (red dots correspond to the baseline, black dots to the LLM-generated function output). An $\times$ indicates cases where the generated function was either not runnable (scores 1--2) or did not return a properly structured output (score 3); hence, no comparison was possible for that seed. Models that only generated such functions are not shown. Identical score plots across different seeds indicate that the respective model produced functions with identical algorithmic processes and internal seeding behavior for two or more seeds.} 
    \label{tab:exp2_micomp}

    {
    \setlength{\tabcolsep}{3pt}    
    \begin{tabular}{lrcrcrcrcrcrc} 
        \toprule
        & \multicolumn{12}{l}{\textbf{Run/Seed}} \\
        \cmidrule(r){2-13}
        & \multicolumn{2}{l}{\textbf{1/0862}}
        & \multicolumn{2}{l}{\textbf{2/1924}}
        & \multicolumn{2}{l}{\textbf{3/2090}}
        & \multicolumn{2}{l}{\textbf{4/4745}}
        & \multicolumn{2}{l}{\textbf{5/5417}}
        & \multicolumn{2}{l}{\textbf{6/7259}}\\
        \cmidrule(r){2-3}
        \cmidrule(r){4-5}
        \cmidrule(r){6-7}
        \cmidrule(r){8-9}
        \cmidrule(r){10-11}
        \cmidrule(r){12-13}
        \textbf{Model}
        & \multicolumn{1}{c}{$p$-val.} & \multicolumn{1}{c}{sp}   
        & \multicolumn{1}{c}{$p$-val.} & \multicolumn{1}{c}{sp}   
        & \multicolumn{1}{c}{$p$-val.} & \multicolumn{1}{c}{sp}   
        & \multicolumn{1}{c}{$p$-val.} & \multicolumn{1}{c}{sp}   
        & \multicolumn{1}{c}{$p$-val.} & \multicolumn{1}{c}{sp}   
        & \multicolumn{1}{c}{$p$-val.} & \multicolumn{1}{c}{sp}   
        \\
        \midrule

\texttt{claude-3.7-sonnet}
 & 0.398 & \raisebox{-.5\height}{\resizebox {1.2cm} {1.2cm} {\begin{tikzpicture}[scale=6] \path (-1.2,-1.2) (1.2,1.2);\draw[very thin,color=gray] (0,1.1)--(0,-1.1); \draw[very thin,color=gray] (1.1,0)--(-1.1,0); \path plot[mark=square*,mark options={color=red},mark size=0.8pt] coordinates { (-0.499,0.262) (-0.591,-0.592) (0.217,0.054) (-0.221,0.104) (-0.364,-0.006) (-0.176,-0.410) (-0.562,-0.191) (-0.090,-0.449) (0.177,0.413) (-0.386,0.195) (-0.065,-0.306) (-0.179,0.141) (-0.154,-0.708) (0.462,-0.326) (-0.146,-0.707) (0.302,-0.285) (-0.216,-0.052) (0.141,0.156) (-0.359,0.403) (0.485,-0.120) (-0.381,-0.145) (-0.153,0.418) (-0.029,0.072) (0.012,0.162) (0.026,0.457) (0.071,0.039) (0.436,-0.243) (0.284,-0.156) (0.003,0.643) (0.516,-0.190)};  \path plot[mark=diamond*,mark options={color=black},mark size=1pt] coordinates { (0.297,0.039) (0.528,0.506) (0.408,-0.034) (0.685,-0.211) (-0.076,-0.094) (-0.669,0.378) (-0.363,-0.089) (0.467,0.208) (-0.225,0.008) (0.098,0.075) (-0.248,0.002) (-0.030,-0.073) (0.550,0.332) (-0.644,0.228) (-0.562,0.025) (-0.033,-0.063) (0.039,0.170) (0.448,0.026) (0.454,0.164) (1.000,-0.104) (0.388,0.216) (-0.456,-0.099) (-0.031,-0.175) (0.065,0.184) (0.069,-0.428) (-0.217,0.314) (-0.106,-0.228) (-0.023,-0.118) (-0.069,-0.096) (-0.306,0.303)};  \end{tikzpicture}}}
 & 0.398 & \raisebox{-.5\height}{\resizebox {1.2cm} {1.2cm} {\begin{tikzpicture}[scale=6] \path (-1.2,-1.2) (1.2,1.2);\draw[very thin,color=gray] (0,1.1)--(0,-1.1); \draw[very thin,color=gray] (1.1,0)--(-1.1,0); \path plot[mark=square*,mark options={color=red},mark size=0.8pt] coordinates { (-0.499,0.262) (-0.591,-0.592) (0.217,0.054) (-0.221,0.104) (-0.364,-0.006) (-0.176,-0.410) (-0.562,-0.191) (-0.090,-0.449) (0.177,0.413) (-0.386,0.195) (-0.065,-0.306) (-0.179,0.141) (-0.154,-0.708) (0.462,-0.326) (-0.146,-0.707) (0.302,-0.285) (-0.216,-0.052) (0.141,0.156) (-0.359,0.403) (0.485,-0.120) (-0.381,-0.145) (-0.153,0.418) (-0.029,0.072) (0.012,0.162) (0.026,0.457) (0.071,0.039) (0.436,-0.243) (0.284,-0.156) (0.003,0.643) (0.516,-0.190)};  \path plot[mark=diamond*,mark options={color=black},mark size=1pt] coordinates { (0.297,0.039) (0.528,0.506) (0.408,-0.034) (0.685,-0.211) (-0.076,-0.094) (-0.669,0.378) (-0.363,-0.089) (0.467,0.208) (-0.225,0.008) (0.098,0.075) (-0.248,0.002) (-0.030,-0.073) (0.550,0.332) (-0.644,0.228) (-0.562,0.025) (-0.033,-0.063) (0.039,0.170) (0.448,0.026) (0.454,0.164) (1.000,-0.104) (0.388,0.216) (-0.456,-0.099) (-0.031,-0.175) (0.065,0.184) (0.069,-0.428) (-0.217,0.314) (-0.106,-0.228) (-0.023,-0.118) (-0.069,-0.096) (-0.306,0.303)};  \end{tikzpicture}}}
 & 0.398 & \raisebox{-.5\height}{\resizebox {1.2cm} {1.2cm} {\begin{tikzpicture}[scale=6] \path (-1.2,-1.2) (1.2,1.2);\draw[very thin,color=gray] (0,1.1)--(0,-1.1); \draw[very thin,color=gray] (1.1,0)--(-1.1,0); \path plot[mark=square*,mark options={color=red},mark size=0.8pt] coordinates { (-0.499,0.262) (-0.591,-0.592) (0.217,0.054) (-0.221,0.104) (-0.364,-0.006) (-0.176,-0.410) (-0.562,-0.191) (-0.090,-0.449) (0.177,0.413) (-0.386,0.195) (-0.065,-0.306) (-0.179,0.141) (-0.154,-0.708) (0.462,-0.326) (-0.146,-0.707) (0.302,-0.285) (-0.216,-0.052) (0.141,0.156) (-0.359,0.403) (0.485,-0.120) (-0.381,-0.145) (-0.153,0.418) (-0.029,0.072) (0.012,0.162) (0.026,0.457) (0.071,0.039) (0.436,-0.243) (0.284,-0.156) (0.003,0.643) (0.516,-0.190)};  \path plot[mark=diamond*,mark options={color=black},mark size=1pt] coordinates { (0.297,0.039) (0.528,0.506) (0.408,-0.034) (0.685,-0.211) (-0.076,-0.094) (-0.669,0.378) (-0.363,-0.089) (0.467,0.208) (-0.225,0.008) (0.098,0.075) (-0.248,0.002) (-0.030,-0.073) (0.550,0.332) (-0.644,0.228) (-0.562,0.025) (-0.033,-0.063) (0.039,0.170) (0.448,0.026) (0.454,0.164) (1.000,-0.104) (0.388,0.216) (-0.456,-0.099) (-0.031,-0.175) (0.065,0.184) (0.069,-0.428) (-0.217,0.314) (-0.106,-0.228) (-0.023,-0.118) (-0.069,-0.096) (-0.306,0.303)};  \end{tikzpicture}}}
 & 0.398 & \raisebox{-.5\height}{\resizebox {1.2cm} {1.2cm} {\begin{tikzpicture}[scale=6] \path (-1.2,-1.2) (1.2,1.2);\draw[very thin,color=gray] (0,1.1)--(0,-1.1); \draw[very thin,color=gray] (1.1,0)--(-1.1,0); \path plot[mark=square*,mark options={color=red},mark size=0.8pt] coordinates { (-0.499,0.262) (-0.591,-0.592) (0.217,0.054) (-0.221,0.104) (-0.364,-0.006) (-0.176,-0.410) (-0.562,-0.191) (-0.090,-0.449) (0.177,0.413) (-0.386,0.195) (-0.065,-0.306) (-0.179,0.141) (-0.154,-0.708) (0.462,-0.326) (-0.146,-0.707) (0.302,-0.285) (-0.216,-0.052) (0.141,0.156) (-0.359,0.403) (0.485,-0.120) (-0.381,-0.145) (-0.153,0.418) (-0.029,0.072) (0.012,0.162) (0.026,0.457) (0.071,0.039) (0.436,-0.243) (0.284,-0.156) (0.003,0.643) (0.516,-0.190)};  \path plot[mark=diamond*,mark options={color=black},mark size=1pt] coordinates { (0.297,0.039) (0.528,0.506) (0.408,-0.034) (0.685,-0.211) (-0.076,-0.094) (-0.669,0.378) (-0.363,-0.089) (0.467,0.208) (-0.225,0.008) (0.098,0.075) (-0.248,0.002) (-0.030,-0.073) (0.550,0.332) (-0.644,0.228) (-0.562,0.025) (-0.033,-0.063) (0.039,0.170) (0.448,0.026) (0.454,0.164) (1.000,-0.104) (0.388,0.216) (-0.456,-0.099) (-0.031,-0.175) (0.065,0.184) (0.069,-0.428) (-0.217,0.314) (-0.106,-0.228) (-0.023,-0.118) (-0.069,-0.096) (-0.306,0.303)};  \end{tikzpicture}}}
 & \multicolumn{1}{c}{$\times$} & \multicolumn{1}{c}{$\times$} & \multicolumn{1}{c}{$\times$} & \multicolumn{1}{c}{$\times$}\\

\texttt{deepseek-r1}
 & 0.256 & \raisebox{-.5\height}{\resizebox {1.2cm} {1.2cm} {\begin{tikzpicture}[scale=6] \path (-1.2,-1.2) (1.2,1.2);\draw[very thin,color=gray] (0,1.1)--(0,-1.1); \draw[very thin,color=gray] (1.1,0)--(-1.1,0); \path plot[mark=square*,mark options={color=red},mark size=0.8pt] coordinates { (-0.345,0.414) (-0.331,-0.476) (0.202,0.007) (-0.071,0.233) (-0.245,0.082) (-0.068,-0.414) (-0.362,-0.042) (0.017,-0.476) (0.223,0.382) (-0.261,0.109) (0.078,-0.276) (-0.060,0.210) (0.017,-0.704) (0.581,-0.267) (0.016,-0.739) (0.398,-0.214) (-0.139,-0.145) (0.224,0.277) (-0.218,0.475) (0.593,0.045) (-0.296,-0.223) (0.018,0.655) (0.064,0.115) (0.091,0.215) (0.117,0.669) (0.129,0.112) (0.472,-0.344) (0.454,-0.069) (0.031,0.548) (0.607,-0.122)};  \path plot[mark=diamond*,mark options={color=black},mark size=1pt] coordinates { (0.384,-0.033) (0.166,0.085) (-0.726,-0.058) (-0.661,0.052) (-0.408,0.069) (-0.353,0.218) (0.040,-0.019) (0.338,0.255) (-0.450,-0.240) (0.535,0.164) (0.503,0.057) (-0.837,-0.119) (0.843,-0.067) (0.462,0.121) (0.465,0.078) (-0.217,0.019) (-0.085,-0.015) (-0.319,-0.160) (0.211,-0.171) (-0.533,-0.128) (-0.278,0.245) (-1.000,0.035) (-0.398,-0.260) (-0.334,-0.238) (-0.379,0.303) (-0.130,0.034) (0.302,-0.201) (-0.073,0.194) (0.817,0.011) (0.184,-0.269)};  \end{tikzpicture}}}
 & \multicolumn{1}{c}{$\times$} & \multicolumn{1}{c}{$\times$} & 0.256 & \raisebox{-.5\height}{\resizebox {1.2cm} {1.2cm} {\begin{tikzpicture}[scale=6] \path (-1.2,-1.2) (1.2,1.2);\draw[very thin,color=gray] (0,1.1)--(0,-1.1); \draw[very thin,color=gray] (1.1,0)--(-1.1,0); \path plot[mark=square*,mark options={color=red},mark size=0.8pt] coordinates { (-0.345,0.414) (-0.331,-0.476) (0.202,0.007) (-0.071,0.233) (-0.245,0.082) (-0.068,-0.414) (-0.362,-0.042) (0.017,-0.476) (0.223,0.382) (-0.261,0.109) (0.078,-0.276) (-0.060,0.210) (0.017,-0.704) (0.581,-0.267) (0.016,-0.739) (0.398,-0.214) (-0.139,-0.145) (0.224,0.277) (-0.218,0.475) (0.593,0.045) (-0.296,-0.223) (0.018,0.655) (0.064,0.115) (0.091,0.215) (0.117,0.669) (0.129,0.112) (0.472,-0.344) (0.454,-0.069) (0.031,0.548) (0.607,-0.122)};  \path plot[mark=diamond*,mark options={color=black},mark size=1pt] coordinates { (0.384,-0.033) (0.166,0.085) (-0.726,-0.058) (-0.661,0.052) (-0.408,0.069) (-0.353,0.218) (0.040,-0.019) (0.338,0.255) (-0.450,-0.240) (0.535,0.164) (0.503,0.057) (-0.837,-0.119) (0.843,-0.067) (0.462,0.121) (0.465,0.078) (-0.217,0.019) (-0.085,-0.015) (-0.319,-0.160) (0.211,-0.171) (-0.533,-0.128) (-0.278,0.245) (-1.000,0.035) (-0.398,-0.260) (-0.334,-0.238) (-0.379,0.303) (-0.130,0.034) (0.302,-0.201) (-0.073,0.194) (0.817,0.011) (0.184,-0.269)};  \end{tikzpicture}}}
 & 0.256 & \raisebox{-.5\height}{\resizebox {1.2cm} {1.2cm} {\begin{tikzpicture}[scale=6] \path (-1.2,-1.2) (1.2,1.2);\draw[very thin,color=gray] (0,1.1)--(0,-1.1); \draw[very thin,color=gray] (1.1,0)--(-1.1,0); \path plot[mark=square*,mark options={color=red},mark size=0.8pt] coordinates { (-0.346,0.411) (-0.327,-0.481) (0.201,0.012) (-0.074,0.232) (-0.249,0.080) (-0.071,-0.415) (-0.362,-0.047) (0.020,-0.471) (0.220,0.386) (-0.260,0.107) (0.078,-0.277) (-0.064,0.210) (0.015,-0.705) (0.578,-0.262) (0.021,-0.736) (0.398,-0.213) (-0.139,-0.141) (0.224,0.281) (-0.221,0.474) (0.590,0.047) (-0.293,-0.219) (0.013,0.649) (0.064,0.118) (0.089,0.216) (0.112,0.668) (0.131,0.118) (0.473,-0.339) (0.455,-0.064) (0.029,0.555) (0.607,-0.119)};  \path plot[mark=diamond*,mark options={color=black},mark size=1pt] coordinates { (0.384,-0.030) (0.169,0.090) (-0.722,-0.058) (-0.663,0.041) (-0.403,0.068) (-0.353,0.225) (0.039,-0.022) (0.337,0.257) (-0.443,-0.231) (0.532,0.161) (0.499,0.059) (-0.835,-0.117) (0.839,-0.078) (0.466,0.135) (0.459,0.081) (-0.219,0.014) (-0.083,-0.018) (-0.317,-0.162) (0.207,-0.185) (-0.533,-0.126) (-0.273,0.248) (-1.000,0.030) (-0.397,-0.261) (-0.329,-0.236) (-0.383,0.291) (-0.136,0.021) (0.302,-0.205) (-0.071,0.194) (0.816,0.009) (0.198,-0.271)};  \end{tikzpicture}}}
 & 0.307 & \raisebox{-.5\height}{\resizebox {1.2cm} {1.2cm} {\begin{tikzpicture}[scale=6] \path (-1.2,-1.2) (1.2,1.2);\draw[very thin,color=gray] (0,1.1)--(0,-1.1); \draw[very thin,color=gray] (1.1,0)--(-1.1,0); \path plot[mark=square*,mark options={color=red},mark size=0.8pt] coordinates { (-0.545,0.064) (-0.090,0.712) (0.327,-0.312) (-0.312,-0.036) (-0.272,0.129) (0.334,0.150) (-0.342,0.417) (0.521,0.252) (-0.028,-0.601) (-0.341,0.439) (0.210,0.329) (-0.253,0.107) (0.416,0.531) (0.607,-0.268) (0.418,0.485) (0.480,-0.185) (0.121,0.114) (0.075,-0.363) (-0.458,0.111) (0.363,-0.587) (0.028,0.437) (-0.615,-0.341) (0.004,0.087) (-0.035,-0.236) (-0.379,-0.497) (0.179,-0.309) (0.680,-0.232) (0.301,-0.243) (-0.344,-0.545) (0.485,-0.393)};  \path plot[mark=diamond*,mark options={color=black},mark size=1pt] coordinates { (0.244,-0.602) (-0.317,0.350) (0.204,0.341) (-0.501,0.087) (-0.090,0.070) (0.084,-0.203) (-0.003,-0.107) (0.287,-0.015) (-0.364,0.173) (-0.160,-0.567) (0.227,0.051) (0.059,0.137) (-0.034,0.191) (0.127,0.275) (0.318,0.536) (-0.107,0.339) (-0.004,-0.029) (-0.350,-0.357) (-0.497,-0.122) (-0.301,-0.122) (-0.181,0.175) (-1.000,-0.113) (0.239,-0.364) (0.140,0.199) (0.128,0.175) (0.363,0.584) (-0.046,0.173) (0.175,-0.281) (-0.762,0.004) (0.586,-0.197)};  \end{tikzpicture}}}
 & \multicolumn{1}{c}{$\times$} & \multicolumn{1}{c}{$\times$}\\

\texttt{deepseek-v3}
 & \multicolumn{1}{c}{$\times$} & \multicolumn{1}{c}{$\times$} & 0.182 & \raisebox{-.5\height}{\resizebox {1.2cm} {1.2cm} {\begin{tikzpicture}[scale=6] \path (-1.2,-1.2) (1.2,1.2);\draw[very thin,color=gray] (0,1.1)--(0,-1.1); \draw[very thin,color=gray] (1.1,0)--(-1.1,0); \path plot[mark=square*,mark options={color=red},mark size=0.8pt] coordinates { (-0.463,0.323) (-0.479,-0.644) (0.387,0.127) (-0.152,0.173) (-0.316,0.040) (0.057,-0.392) (-0.485,-0.155) (0.103,-0.457) (0.287,0.512) (-0.268,-0.007) (0.097,-0.310) (-0.053,0.159) (0.056,-0.753) (0.648,-0.182) (0.068,-0.786) (0.453,-0.150) (-0.057,-0.108) (0.214,0.341) (-0.307,0.399) (0.614,0.122) (-0.231,-0.252) (-0.200,0.582) (0.071,0.099) (0.140,0.246) (0.013,0.698) (0.186,0.192) (0.601,-0.234) (0.388,-0.048) (0.082,0.600) (0.695,-0.034)};  \path plot[mark=diamond*,mark options={color=black},mark size=1pt] coordinates { (-0.402,0.044) (0.303,0.101) (-0.729,0.114) (0.159,-0.081) (0.461,-0.213) (0.299,0.123) (-0.273,0.175) (-0.081,-0.314) (0.155,0.126) (0.644,0.016) (-0.378,0.013) (-0.186,-0.017) (-0.430,0.019) (-0.038,-0.268) (-0.203,-0.048) (0.136,-0.284) (-0.245,-0.007) (-0.155,0.024) (-0.488,-0.033) (-0.694,0.119) (0.014,0.228) (0.198,0.021) (-0.075,-0.209) (-0.100,-0.160) (-0.073,-0.057) (-0.101,-0.019) (0.389,0.162) (0.878,0.331) (-1.000,0.112) (-0.137,-0.119)};  \end{tikzpicture}}}
 & \multicolumn{1}{c}{$\times$} & \multicolumn{1}{c}{$\times$} & \multicolumn{1}{c}{$\times$} & \multicolumn{1}{c}{$\times$} & 0.182 & \raisebox{-.5\height}{\resizebox {1.2cm} {1.2cm} {\begin{tikzpicture}[scale=6] \path (-1.2,-1.2) (1.2,1.2);\draw[very thin,color=gray] (0,1.1)--(0,-1.1); \draw[very thin,color=gray] (1.1,0)--(-1.1,0); \path plot[mark=square*,mark options={color=red},mark size=0.8pt] coordinates { (-0.345,0.414) (-0.331,-0.476) (0.202,0.007) (-0.071,0.233) (-0.245,0.082) (-0.068,-0.414) (-0.362,-0.042) (0.017,-0.476) (0.223,0.382) (-0.261,0.109) (0.078,-0.276) (-0.060,0.210) (0.017,-0.704) (0.581,-0.267) (0.016,-0.739) (0.398,-0.214) (-0.139,-0.145) (0.224,0.277) (-0.218,0.475) (0.593,0.045) (-0.296,-0.223) (0.018,0.655) (0.064,0.115) (0.091,0.215) (0.117,0.669) (0.129,0.112) (0.472,-0.344) (0.454,-0.069) (0.031,0.548) (0.607,-0.122)};  \path plot[mark=diamond*,mark options={color=black},mark size=1pt] coordinates { (0.384,-0.033) (0.166,0.085) (-0.726,-0.058) (-0.661,0.052) (-0.408,0.069) (-0.353,0.218) (0.040,-0.019) (0.338,0.255) (-0.450,-0.240) (0.535,0.164) (0.503,0.057) (-0.837,-0.119) (0.843,-0.067) (0.462,0.121) (0.465,0.078) (-0.217,0.019) (-0.085,-0.015) (-0.319,-0.160) (0.211,-0.171) (-0.533,-0.128) (-0.278,0.245) (-1.000,0.035) (-0.398,-0.260) (-0.334,-0.238) (-0.379,0.303) (-0.130,0.034) (0.302,-0.201) (-0.073,0.194) (0.817,0.011) (0.184,-0.269)};  \end{tikzpicture}}}
 & \multicolumn{1}{c}{$\times$} & \multicolumn{1}{c}{$\times$}\\

\texttt{gpt-4.1}
 & 0.224 & \raisebox{-.5\height}{\resizebox {1.2cm} {1.2cm} {\begin{tikzpicture}[scale=6] \path (-1.2,-1.2) (1.2,1.2);\draw[very thin,color=gray] (0,1.1)--(0,-1.1); \draw[very thin,color=gray] (1.1,0)--(-1.1,0); \path plot[mark=square*,mark options={color=red},mark size=0.8pt] coordinates { (-0.482,0.152) (-0.543,-0.507) (0.066,0.299) (-0.218,0.051) (-0.411,-0.060) (-0.186,-0.281) (-0.577,-0.193) (-0.129,-0.129) (0.047,0.574) (-0.296,0.006) (-0.023,-0.357) (-0.128,-0.044) (-0.136,-0.644) (0.405,-0.070) (-0.142,-0.497) (0.225,-0.084) (-0.198,0.063) (0.061,0.347) (-0.358,0.220) (0.439,0.145) (-0.434,-0.037) (-0.166,0.241) (-0.032,0.087) (-0.028,0.180) (-0.051,0.352) (-0.035,0.269) (0.344,0.037) (0.277,0.064) (-0.032,0.637) (0.465,0.109)};  \path plot[mark=diamond*,mark options={color=black},mark size=1pt] coordinates { (0.543,-0.234) (0.009,-0.362) (-0.416,-0.004) (0.442,0.086) (0.137,0.305) (0.274,-0.479) (0.255,-0.553) (-0.225,0.482) (-0.064,-0.309) (-0.433,0.369) (0.317,-0.035) (0.385,-0.592) (0.874,-0.041) (0.793,0.068) (1.000,0.325) (-0.087,0.180) (0.221,0.162) (0.635,0.149) (-0.062,-0.181) (0.061,-0.066) (0.016,0.305) (-0.177,-0.158) (0.048,-0.354) (-0.000,-0.384) (-0.094,-0.268) (-0.633,0.009) (0.096,0.436) (-0.974,0.245) (0.167,-0.082) (-0.834,0.051)};  \end{tikzpicture}}}
 & 0.439 & \raisebox{-.5\height}{\resizebox {1.2cm} {1.2cm} {\begin{tikzpicture}[scale=6] \path (-1.2,-1.2) (1.2,1.2);\draw[very thin,color=gray] (0,1.1)--(0,-1.1); \draw[very thin,color=gray] (1.1,0)--(-1.1,0); \path plot[mark=square*,mark options={color=red},mark size=0.8pt] coordinates { (0.402,-0.688) (0.047,-0.352) (-0.195,0.273) (0.066,-0.400) (0.098,-0.400) (-0.566,0.074) (0.234,-0.525) (-0.220,0.127) (0.447,0.211) (0.099,-0.567) (-0.388,-0.152) (0.014,-0.400) (-0.584,0.089) (-0.060,0.630) (-0.599,0.107) (-0.085,0.457) (-0.252,-0.044) (0.207,0.099) (0.348,-0.863) (0.374,0.825) (-0.332,-0.482) (0.976,-0.169) (-0.126,-0.209) (0.186,-0.042) (0.702,-0.308) (0.187,0.074) (-0.010,0.682) (0.129,0.480) (0.594,-0.010) (-0.057,0.763)};  \path plot[mark=diamond*,mark options={color=black},mark size=1pt] coordinates { (0.627,0.727) (-0.850,0.018) (0.046,0.188) (-0.200,-0.580) (0.140,-0.082) (-0.320,-0.258) (-0.550,0.089) (-0.260,-0.029) (0.157,-0.516) (0.673,-0.564) (0.139,-0.237) (0.023,0.614) (0.313,-0.056) (-0.411,-0.020) (1.000,0.528) (0.744,0.198) (-0.022,0.521) (-0.591,0.157) (-0.692,-0.031) (0.607,0.144) (-0.418,0.501) (0.105,0.296) (-0.195,-0.135) (-0.474,0.154) (-0.154,-0.437) (0.326,-0.486) (-0.358,-0.080) (-0.571,-0.192) (0.243,0.469) (-0.708,-0.180)};  \end{tikzpicture}}}
 & 0.224 & \raisebox{-.5\height}{\resizebox {1.2cm} {1.2cm} {\begin{tikzpicture}[scale=6] \path (-1.2,-1.2) (1.2,1.2);\draw[very thin,color=gray] (0,1.1)--(0,-1.1); \draw[very thin,color=gray] (1.1,0)--(-1.1,0); \path plot[mark=square*,mark options={color=red},mark size=0.8pt] coordinates { (-0.482,0.152) (-0.543,-0.507) (0.066,0.299) (-0.218,0.051) (-0.411,-0.060) (-0.186,-0.281) (-0.577,-0.193) (-0.129,-0.129) (0.047,0.574) (-0.296,0.006) (-0.023,-0.357) (-0.128,-0.044) (-0.136,-0.644) (0.405,-0.070) (-0.142,-0.497) (0.225,-0.084) (-0.198,0.063) (0.061,0.347) (-0.358,0.220) (0.439,0.145) (-0.434,-0.037) (-0.166,0.241) (-0.032,0.087) (-0.028,0.180) (-0.051,0.352) (-0.035,0.269) (0.344,0.037) (0.277,0.064) (-0.032,0.637) (0.465,0.109)};  \path plot[mark=diamond*,mark options={color=black},mark size=1pt] coordinates { (0.543,-0.234) (0.009,-0.362) (-0.416,-0.004) (0.442,0.086) (0.137,0.305) (0.274,-0.479) (0.255,-0.553) (-0.225,0.482) (-0.064,-0.309) (-0.433,0.369) (0.317,-0.035) (0.385,-0.592) (0.874,-0.041) (0.793,0.068) (1.000,0.325) (-0.087,0.180) (0.221,0.162) (0.635,0.149) (-0.062,-0.181) (0.061,-0.066) (0.016,0.305) (-0.177,-0.158) (0.048,-0.354) (-0.000,-0.384) (-0.094,-0.268) (-0.633,0.009) (0.096,0.436) (-0.974,0.245) (0.167,-0.082) (-0.834,0.051)};  \end{tikzpicture}}}
 & 0.395 & \raisebox{-.5\height}{\resizebox {1.2cm} {1.2cm} {\begin{tikzpicture}[scale=6] \path (-1.2,-1.2) (1.2,1.2);\draw[very thin,color=gray] (0,1.1)--(0,-1.1); \draw[very thin,color=gray] (1.1,0)--(-1.1,0); \path plot[mark=square*,mark options={color=red},mark size=0.8pt] coordinates { (-0.384,0.367) (-0.422,-0.697) (0.294,0.085) (-0.098,0.227) (-0.266,0.075) (-0.008,-0.350) (-0.420,-0.156) (0.053,-0.465) (0.243,0.427) (-0.308,0.053) (0.040,-0.283) (-0.057,0.237) (0.006,-0.755) (0.614,-0.234) (0.019,-0.760) (0.428,-0.206) (-0.125,-0.041) (0.257,0.364) (-0.274,0.473) (0.627,0.025) (-0.282,-0.195) (-0.071,0.491) (0.084,0.219) (0.096,0.230) (0.071,0.656) (0.173,0.197) (0.575,-0.294) (0.421,-0.095) (0.049,0.589) (0.643,-0.090)};  \path plot[mark=diamond*,mark options={color=black},mark size=1pt] coordinates { (-0.279,-0.060) (-0.953,0.148) (-0.207,-0.246) (0.448,0.104) (0.011,-0.293) (-0.829,0.112) (-0.553,-0.035) (0.006,0.011) (0.865,-0.080) (-0.547,0.141) (-0.155,-0.127) (0.198,0.201) (0.080,-0.053) (-0.042,-0.078) (0.728,0.081) (0.039,0.002) (-0.007,-0.001) (-0.164,-0.112) (-0.681,0.050) (-0.766,0.060) (-0.236,0.035) (-0.264,-0.031) (0.542,-0.098) (1.000,0.230) (-0.365,-0.001) (0.155,-0.082) (-0.294,-0.125) (-0.293,-0.109) (0.258,0.076) (0.328,0.184)};  \end{tikzpicture}}}
 & 0.633 & \raisebox{-.5\height}{\resizebox {1.2cm} {1.2cm} {\begin{tikzpicture}[scale=6] \path (-1.2,-1.2) (1.2,1.2);\draw[very thin,color=gray] (0,1.1)--(0,-1.1); \draw[very thin,color=gray] (1.1,0)--(-1.1,0); \path plot[mark=square*,mark options={color=red},mark size=0.8pt] coordinates { (-0.741,0.027) (-0.295,-0.733) (0.387,0.309) (-0.113,-0.087) (-0.207,-0.109) (0.524,-0.360) (-0.389,-0.451) (0.239,-0.299) (0.114,0.598) (-0.540,-0.315) (0.074,-0.351) (-0.357,-0.142) (0.417,-0.673) (0.643,0.266) (0.395,-0.587) (0.484,0.154) (0.136,-0.185) (0.040,0.462) (-0.428,-0.023) (0.431,0.506) (0.040,-0.521) (-0.611,0.307) (-0.043,-0.009) (0.005,0.202) (-0.377,0.532) (0.085,0.335) (0.511,0.194) (0.322,0.258) (-0.296,0.597) (0.606,0.442)};  \path plot[mark=diamond*,mark options={color=black},mark size=1pt] coordinates { (0.220,0.257) (-0.057,-0.176) (0.182,-0.197) (0.279,-0.294) (-0.920,-0.379) (0.018,0.448) (-0.008,-0.189) (-0.058,-0.143) (-0.617,-0.489) (-0.204,-0.230) (0.066,0.029) (0.348,0.148) (0.098,-0.355) (0.402,-0.284) (-0.363,0.258) (-0.473,0.073) (0.003,0.539) (0.179,0.482) (0.292,0.081) (0.393,-0.254) (0.232,-0.176) (0.186,-0.377) (-1.000,0.338) (-0.395,0.575) (-0.287,0.528) (0.510,-0.639) (0.579,0.039) (0.386,0.378) (-0.160,-0.192) (-0.886,-0.146)};  \end{tikzpicture}}}
 & 0.633 & \raisebox{-.5\height}{\resizebox {1.2cm} {1.2cm} {\begin{tikzpicture}[scale=6] \path (-1.2,-1.2) (1.2,1.2);\draw[very thin,color=gray] (0,1.1)--(0,-1.1); \draw[very thin,color=gray] (1.1,0)--(-1.1,0); \path plot[mark=square*,mark options={color=red},mark size=0.8pt] coordinates { (0.454,-0.341) (0.591,0.546) (-0.229,-0.224) (0.159,-0.186) (0.390,-0.053) (0.078,0.328) (0.567,0.106) (0.003,0.302) (-0.178,-0.564) (0.359,-0.006) (0.008,0.375) (0.072,-0.073) (0.109,0.723) (-0.501,0.188) (0.035,0.683) (-0.279,0.134) (0.096,0.057) (-0.208,-0.387) (0.349,-0.380) (-0.491,-0.149) (0.284,0.157) (0.246,-0.569) (-0.145,-0.084) (-0.013,-0.226) (0.071,-0.612) (-0.111,-0.295) (-0.496,0.089) (-0.316,-0.024) (-0.006,-0.687) (-0.594,0.014)};  \path plot[mark=diamond*,mark options={color=black},mark size=1pt] coordinates { (-0.206,0.281) (-0.190,0.303) (-0.279,0.290) (-0.080,-0.161) (-0.302,-0.086) (-0.665,-0.169) (-0.314,0.091) (0.460,0.069) (0.044,-0.216) (0.571,0.177) (0.503,0.179) (0.370,-0.202) (0.068,0.032) (0.968,-0.139) (0.959,-0.157) (0.017,0.221) (0.300,0.024) (-0.236,-0.304) (-0.114,0.223) (1.000,-0.231) (-0.847,-0.606) (-0.071,0.342) (0.199,0.137) (-0.962,0.145) (0.372,0.112) (0.031,0.205) (-0.707,0.265) (-0.278,0.054) (-0.681,-0.036) (-0.235,0.314)};  \end{tikzpicture}}}
\\

\texttt{gpt-4o}
 & 0.174 & \raisebox{-.5\height}{\resizebox {1.2cm} {1.2cm} {\begin{tikzpicture}[scale=6] \path (-1.2,-1.2) (1.2,1.2);\draw[very thin,color=gray] (0,1.1)--(0,-1.1); \draw[very thin,color=gray] (1.1,0)--(-1.1,0); \path plot[mark=square*,mark options={color=red},mark size=0.8pt] coordinates { (-0.346,0.190) (-0.409,0.466) (0.027,-0.424) (-0.129,0.179) (-0.321,0.175) (-0.180,0.062) (-0.396,0.345) (-0.138,-0.060) (0.022,-0.480) (-0.303,0.225) (-0.012,0.243) (-0.108,0.233) (-0.124,0.292) (0.386,-0.129) (-0.122,0.165) (0.229,-0.122) (-0.196,-0.063) (0.137,-0.229) (-0.237,0.237) (0.398,-0.222) (-0.414,0.084) (-0.074,0.164) (0.003,0.092) (-0.028,-0.061) (0.037,-0.011) (0.038,-0.197) (0.272,-0.268) (0.289,-0.102) (-0.083,-0.317) (0.431,-0.304)};  \path plot[mark=diamond*,mark options={color=black},mark size=1pt] coordinates { (0.326,-0.390) (0.054,0.208) (0.537,-0.096) (0.552,-0.118) (0.887,0.356) (-0.183,0.077) (0.572,-0.331) (-0.089,0.051) (0.413,0.051) (0.470,-0.141) (-0.547,-0.345) (0.363,0.076) (0.078,0.361) (-0.403,-0.604) (0.017,0.363) (-0.044,0.132) (0.021,-0.443) (-0.019,-0.033) (-0.048,0.089) (0.019,0.217) (0.139,0.187) (0.111,0.428) (0.278,0.284) (-1.000,-0.605) (-0.617,0.141) (-0.063,0.192) (0.119,0.173) (-0.890,-0.475) (-0.249,0.348) (0.545,-0.319)};  \end{tikzpicture}}}
 & 0.174 & \raisebox{-.5\height}{\resizebox {1.2cm} {1.2cm} {\begin{tikzpicture}[scale=6] \path (-1.2,-1.2) (1.2,1.2);\draw[very thin,color=gray] (0,1.1)--(0,-1.1); \draw[very thin,color=gray] (1.1,0)--(-1.1,0); \path plot[mark=square*,mark options={color=red},mark size=0.8pt] coordinates { (-0.490,0.231) (-0.493,-0.673) (0.095,0.150) (-0.266,0.132) (-0.424,-0.027) (-0.158,-0.329) (-0.525,-0.269) (-0.075,-0.315) (0.053,0.371) (-0.341,0.057) (-0.013,-0.286) (-0.151,0.112) (-0.062,-0.690) (0.436,-0.216) (-0.089,-0.596) (0.245,-0.186) (-0.160,-0.031) (0.085,0.314) (-0.391,0.308) (0.418,0.016) (-0.409,-0.120) (-0.233,0.258) (-0.045,0.197) (-0.058,0.151) (-0.088,0.457) (0.007,0.213) (0.355,-0.147) (0.288,-0.054) (-0.099,0.578) (0.450,-0.029)};  \path plot[mark=diamond*,mark options={color=black},mark size=1pt] coordinates { (0.147,0.095) (0.892,0.304) (0.704,0.507) (-0.396,0.596) (-0.071,-0.195) (-1.000,-0.013) (0.037,0.066) (-0.809,0.221) (0.686,0.002) (0.690,-0.129) (-0.100,-0.096) (0.415,-0.106) (0.164,-0.155) (-0.859,-0.028) (0.072,0.168) (-0.193,0.342) (-0.179,-0.233) (-0.104,-0.214) (0.529,0.185) (0.559,-0.308) (0.124,-0.345) (0.038,-0.078) (0.351,-0.016) (-0.323,0.032) (0.177,-0.215) (0.490,-0.149) (0.414,-0.107) (-0.558,-0.001) (-0.210,0.104) (0.455,0.191)};  \end{tikzpicture}}}
 & \multicolumn{1}{c}{$\times$} & \multicolumn{1}{c}{$\times$} & \multicolumn{1}{c}{$\times$} & \multicolumn{1}{c}{$\times$} & \multicolumn{1}{c}{$\times$} & \multicolumn{1}{c}{$\times$} & 0.366 & \raisebox{-.5\height}{\resizebox {1.2cm} {1.2cm} {\begin{tikzpicture}[scale=6] \path (-1.2,-1.2) (1.2,1.2);\draw[very thin,color=gray] (0,1.1)--(0,-1.1); \draw[very thin,color=gray] (1.1,0)--(-1.1,0); \path plot[mark=square*,mark options={color=red},mark size=0.8pt] coordinates { (0.645,-0.197) (0.137,-1.000) (-0.299,0.400) (0.369,-0.092) (0.351,-0.083) (-0.352,-0.205) (0.441,-0.443) (-0.409,-0.429) (-0.195,0.723) (0.259,-0.565) (-0.321,-0.206) (0.057,0.034) (-0.595,-0.500) (-0.703,0.303) (-0.607,-0.683) (-0.402,0.272) (-0.178,-0.120) (0.009,0.306) (0.610,-0.180) (-0.268,0.557) (-0.055,-0.534) (0.720,0.361) (0.100,-0.104) (0.020,0.330) (0.518,0.631) (-0.044,0.117) (-0.673,0.239) (-0.312,0.012) (0.311,0.274) (-0.618,0.429)};  \path plot[mark=diamond*,mark options={color=black},mark size=1pt] coordinates { (-0.161,0.677) (0.054,-0.605) (0.209,0.199) (0.732,-0.133) (-0.639,-0.242) (0.632,0.512) (-0.161,0.225) (-0.440,-0.601) (-0.350,-0.160) (-0.208,0.026) (0.706,-0.274) (0.734,-0.164) (0.612,0.743) (0.285,0.416) (-0.534,-0.049) (-0.077,0.105) (-0.096,-0.635) (0.012,-0.209) (-0.280,-0.010) (0.500,0.107) (0.537,-0.465) (-0.749,0.524) (-0.193,0.319) (0.564,-0.071) (-0.202,0.117) (-0.242,0.372) (0.109,-0.190) (-0.096,-0.483) (0.046,-0.166) (0.180,0.468)};  \end{tikzpicture}}}
\\

\texttt{grok-3-beta}
 & 0.336 & \raisebox{-.5\height}{\resizebox {1.2cm} {1.2cm} {\begin{tikzpicture}[scale=6] \path (-1.2,-1.2) (1.2,1.2);\draw[very thin,color=gray] (0,1.1)--(0,-1.1); \draw[very thin,color=gray] (1.1,0)--(-1.1,0); \path plot[mark=square*,mark options={color=red},mark size=0.8pt] coordinates { (-0.345,0.418) (-0.329,-0.480) (0.212,0.020) (-0.072,0.234) (-0.246,0.081) (-0.066,-0.417) (-0.359,-0.046) (0.031,-0.463) (0.231,0.409) (-0.257,0.103) (0.081,-0.289) (-0.062,0.205) (0.017,-0.709) (0.590,-0.261) (0.024,-0.736) (0.410,-0.214) (-0.130,-0.138) (0.236,0.294) (-0.215,0.470) (0.605,0.052) (-0.287,-0.214) (0.018,0.654) (0.076,0.112) (0.095,0.217) (0.119,0.669) (0.149,0.123) (0.486,-0.332) (0.470,-0.056) (0.039,0.563) (0.619,-0.112)};  \path plot[mark=diamond*,mark options={color=black},mark size=1pt] coordinates { (0.174,0.095) (-0.718,-0.048) (-0.663,0.045) (-0.403,0.066) (-0.344,0.233) (0.042,-0.017) (0.346,0.261) (-0.432,-0.224) (0.551,0.170) (0.504,0.066) (-0.833,-0.107) (0.847,-0.081) (0.479,0.143) (0.475,0.085) (-0.221,0.012) (-0.076,-0.013) (-0.321,-0.163) (0.210,-0.190) (-0.523,-0.115) (-0.267,0.256) (-1.000,0.044) (-0.395,-0.260) (-0.319,-0.237) (-0.387,0.290) (-0.135,0.019) (0.302,-0.216) (-0.062,0.201) (0.822,0.002) (0.197,-0.287) (0.007,-0.183)};  \end{tikzpicture}}}
 & 0.336 & \raisebox{-.5\height}{\resizebox {1.2cm} {1.2cm} {\begin{tikzpicture}[scale=6] \path (-1.2,-1.2) (1.2,1.2);\draw[very thin,color=gray] (0,1.1)--(0,-1.1); \draw[very thin,color=gray] (1.1,0)--(-1.1,0); \path plot[mark=square*,mark options={color=red},mark size=0.8pt] coordinates { (-0.345,0.418) (-0.329,-0.480) (0.212,0.020) (-0.072,0.234) (-0.246,0.081) (-0.066,-0.417) (-0.359,-0.046) (0.031,-0.463) (0.231,0.409) (-0.257,0.103) (0.081,-0.289) (-0.062,0.205) (0.017,-0.709) (0.590,-0.261) (0.024,-0.736) (0.410,-0.214) (-0.130,-0.138) (0.236,0.294) (-0.215,0.470) (0.605,0.052) (-0.287,-0.214) (0.018,0.654) (0.076,0.112) (0.095,0.217) (0.119,0.669) (0.149,0.123) (0.486,-0.332) (0.470,-0.056) (0.039,0.563) (0.619,-0.112)};  \path plot[mark=diamond*,mark options={color=black},mark size=1pt] coordinates { (0.174,0.095) (-0.718,-0.048) (-0.663,0.045) (-0.403,0.066) (-0.344,0.233) (0.042,-0.017) (0.346,0.261) (-0.432,-0.224) (0.551,0.170) (0.504,0.066) (-0.833,-0.107) (0.847,-0.081) (0.479,0.143) (0.475,0.085) (-0.221,0.012) (-0.076,-0.013) (-0.321,-0.163) (0.210,-0.190) (-0.523,-0.115) (-0.267,0.256) (-1.000,0.044) (-0.395,-0.260) (-0.319,-0.237) (-0.387,0.290) (-0.135,0.019) (0.302,-0.216) (-0.062,0.201) (0.822,0.002) (0.197,-0.287) (0.007,-0.183)};  \end{tikzpicture}}}
 & 0.398 & \raisebox{-.5\height}{\resizebox {1.2cm} {1.2cm} {\begin{tikzpicture}[scale=6] \path (-1.2,-1.2) (1.2,1.2);\draw[very thin,color=gray] (0,1.1)--(0,-1.1); \draw[very thin,color=gray] (1.1,0)--(-1.1,0); \path plot[mark=square*,mark options={color=red},mark size=0.8pt] coordinates { (0.651,-0.191) (0.172,-1.000) (-0.301,0.417) (0.377,-0.071) (0.351,-0.075) (-0.362,-0.197) (0.463,-0.440) (-0.411,-0.434) (-0.204,0.731) (0.262,-0.585) (-0.321,-0.227) (0.050,0.017) (-0.585,-0.519) (-0.726,0.302) (-0.600,-0.693) (-0.394,0.277) (-0.192,-0.121) (0.003,0.318) (0.631,-0.181) (-0.274,0.581) (-0.051,-0.532) (0.740,0.387) (0.099,-0.104) (0.016,0.336) (0.513,0.638) (-0.054,0.115) (-0.692,0.232) (-0.323,0.021) (0.295,0.275) (-0.617,0.446)};  \path plot[mark=diamond*,mark options={color=black},mark size=1pt] coordinates { (-0.182,0.687) (0.080,-0.596) (0.224,0.222) (0.726,-0.139) (-0.652,-0.241) (0.650,0.522) (-0.161,0.225) (-0.455,-0.607) (-0.372,-0.184) (-0.233,0.024) (0.715,-0.281) (0.744,-0.144) (0.626,0.767) (0.302,0.438) (-0.527,-0.074) (-0.071,0.101) (-0.108,-0.652) (0.027,-0.238) (-0.281,-0.002) (0.505,0.090) (0.598,-0.437) (-0.757,0.517) (-0.194,0.291) (0.559,-0.063) (-0.221,0.114) (-0.265,0.393) (0.121,-0.194) (-0.125,-0.507) (0.052,-0.213) (0.155,0.456)};  \end{tikzpicture}}}
 & 0.398 & \raisebox{-.5\height}{\resizebox {1.2cm} {1.2cm} {\begin{tikzpicture}[scale=6] \path (-1.2,-1.2) (1.2,1.2);\draw[very thin,color=gray] (0,1.1)--(0,-1.1); \draw[very thin,color=gray] (1.1,0)--(-1.1,0); \path plot[mark=square*,mark options={color=red},mark size=0.8pt] coordinates { (-0.656,0.510) (0.171,0.452) (-0.046,-0.412) (-0.423,0.356) (-0.175,0.432) (0.448,0.078) (-0.084,0.543) (0.376,-0.221) (-0.571,-0.532) (-0.018,0.652) (0.427,0.113) (-0.160,0.463) (0.676,0.197) (0.333,-0.684) (0.683,0.052) (0.265,-0.570) (0.279,0.226) (-0.355,-0.407) (-0.470,0.493) (-0.183,-0.819) (0.209,0.526) (-1.000,0.122) (0.072,0.214) (-0.242,-0.093) (-0.836,-0.128) (-0.159,-0.402) (0.279,-0.801) (0.060,-0.459) (-0.697,0.010) (0.150,-0.828)};  \path plot[mark=diamond*,mark options={color=black},mark size=1pt] coordinates { (-0.947,0.053) (-0.644,0.574) (0.551,0.138) (0.566,-0.038) (0.102,-0.076) (0.688,-0.335) (-0.583,-0.006) (0.509,0.231) (-0.007,0.163) (-0.243,0.368) (0.522,0.169) (-0.666,-0.540) (0.426,0.232) (0.063,0.315) (0.202,0.085) (0.144,-0.458) (0.341,0.397) (-0.188,-0.797) (-0.429,-0.524) (0.313,0.174) (-0.026,0.241) (0.041,0.340) (0.284,-0.574) (0.399,0.071) (-0.151,-0.024) (-0.842,0.060) (0.449,0.086) (0.078,0.168) (-0.019,0.177) (0.715,0.246)};  \end{tikzpicture}}}
 & 0.398 & \raisebox{-.5\height}{\resizebox {1.2cm} {1.2cm} {\begin{tikzpicture}[scale=6] \path (-1.2,-1.2) (1.2,1.2);\draw[very thin,color=gray] (0,1.1)--(0,-1.1); \draw[very thin,color=gray] (1.1,0)--(-1.1,0); \path plot[mark=square*,mark options={color=red},mark size=0.8pt] coordinates { (-0.516,0.266) (-0.592,-0.602) (0.244,0.064) (-0.224,0.105) (-0.368,-0.014) (-0.162,-0.418) (-0.573,-0.196) (-0.080,-0.454) (0.192,0.428) (-0.394,0.191) (-0.061,-0.315) (-0.184,0.142) (-0.138,-0.723) (0.480,-0.328) (-0.138,-0.717) (0.312,-0.282) (-0.218,-0.056) (0.137,0.168) (-0.372,0.403) (0.491,-0.114) (-0.370,-0.150) (-0.162,0.428) (-0.040,0.067) (0.011,0.165) (0.013,0.466) (0.062,0.046) (0.457,-0.236) (0.289,-0.155) (0.009,0.653) (0.526,-0.180)};  \path plot[mark=diamond*,mark options={color=black},mark size=1pt] coordinates { (0.296,0.051) (0.523,0.474) (0.405,-0.062) (0.676,-0.210) (-0.066,-0.086) (-0.669,0.400) (-0.378,-0.103) (0.490,0.224) (-0.238,0.006) (0.090,0.087) (-0.263,-0.003) (-0.038,-0.058) (0.559,0.349) (-0.644,0.233) (-0.568,0.037) (-0.042,-0.056) (0.045,0.140) (0.446,0.017) (0.453,0.177) (1.000,-0.090) (0.386,0.209) (-0.456,-0.119) (-0.043,-0.191) (0.066,0.197) (0.065,-0.442) (-0.216,0.320) (-0.112,-0.218) (-0.024,-0.108) (-0.076,-0.108) (-0.298,0.283)};  \end{tikzpicture}}}
 & \multicolumn{1}{c}{$\times$} & \multicolumn{1}{c}{$\times$}\\

\texttt{mistral-large}
 & 0.582 & \raisebox{-.5\height}{\resizebox {1.2cm} {1.2cm} {\begin{tikzpicture}[scale=6] \path (-1.2,-1.2) (1.2,1.2);\draw[very thin,color=gray] (0,1.1)--(0,-1.1); \draw[very thin,color=gray] (1.1,0)--(-1.1,0); \path plot[mark=square*,mark options={color=red},mark size=0.8pt] coordinates { (0.280,-0.806) (0.947,0.411) (-0.426,-0.046) (0.223,-0.450) (0.283,-0.325) (0.127,0.417) (0.575,-0.320) (0.162,0.531) (-0.684,-0.517) (0.602,-0.406) (0.260,0.205) (0.198,-0.511) (0.563,0.718) (-0.572,0.530) (0.484,0.903) (-0.452,0.411) (-0.021,-0.004) (-0.686,-0.345) (0.436,-0.900) (-0.910,0.293) (0.718,-0.079) (-0.133,-0.811) (-0.065,-0.343) (-0.246,-0.236) (-0.376,-0.902) (-0.433,-0.140) (-0.472,0.661) (-0.382,0.264) (-0.235,-0.561) (-1.000,0.368)};  \path plot[mark=diamond*,mark options={color=black},mark size=1pt] coordinates { (0.121,0.365) (-0.536,0.457) (-0.151,0.097) (0.346,0.514) (0.027,-0.351) (0.415,0.089) (0.030,0.213) (-0.021,0.592) (-0.316,-0.446) (0.289,-0.236) (0.174,-0.151) (0.556,-0.252) (-0.104,-0.012) (-0.474,-0.436) (-0.642,-0.192) (0.661,0.040) (0.395,0.583) (0.242,-0.221) (-0.063,-0.222) (0.137,0.022) (-0.318,0.344) (-0.232,0.430) (0.676,0.137) (-0.640,-0.253) (0.516,-0.350) (0.175,0.049) (-0.530,0.434) (0.155,0.365) (-0.027,0.256) (0.372,0.126)};  \end{tikzpicture}}}
 & 0.582 & \raisebox{-.5\height}{\resizebox {1.2cm} {1.2cm} {\begin{tikzpicture}[scale=6] \path (-1.2,-1.2) (1.2,1.2);\draw[very thin,color=gray] (0,1.1)--(0,-1.1); \draw[very thin,color=gray] (1.1,0)--(-1.1,0); \path plot[mark=square*,mark options={color=red},mark size=0.8pt] coordinates { (-0.669,0.429) (0.179,0.825) (0.130,-0.483) (-0.562,0.252) (-0.259,0.113) (0.472,0.032) (-0.169,0.547) (0.638,0.189) (-0.251,-0.806) (-0.107,0.803) (0.466,0.208) (0.047,0.348) (0.836,0.271) (0.646,-0.651) (0.792,0.463) (0.402,-0.497) (0.284,0.283) (-0.232,-0.371) (-0.912,0.323) (0.238,-0.793) (0.015,0.756) (-0.876,-0.146) (-0.026,0.442) (-0.143,-0.260) (-0.830,-0.678) (-0.049,-0.383) (0.754,-0.546) (0.259,-0.300) (-0.632,-0.245) (0.509,-0.576)};  \path plot[mark=diamond*,mark options={color=black},mark size=1pt] coordinates { (0.321,-0.182) (0.134,0.287) (0.287,0.123) (-0.094,0.423) (-0.234,-0.519) (0.073,0.428) (0.480,-0.720) (-0.492,0.774) (0.803,0.375) (0.062,0.369) (0.358,-0.177) (-0.189,-0.695) (-0.121,0.191) (-0.004,-0.208) (0.108,-0.396) (-0.394,-0.150) (-0.437,-0.008) (0.735,-0.386) (0.233,0.287) (-0.248,0.245) (-0.923,0.028) (-0.054,-0.410) (-0.964,-0.493) (0.589,0.502) (-0.265,0.109) (0.105,0.704) (-0.671,0.034) (-1.000,0.132) (0.161,-0.573) (0.693,0.357)};  \end{tikzpicture}}}
 & 0.296 & \raisebox{-.5\height}{\resizebox {1.2cm} {1.2cm} {\begin{tikzpicture}[scale=6] \path (-1.2,-1.2) (1.2,1.2);\draw[very thin,color=gray] (0,1.1)--(0,-1.1); \draw[very thin,color=gray] (1.1,0)--(-1.1,0); \path plot[mark=square*,mark options={color=red},mark size=0.8pt] coordinates { (-0.345,0.414) (-0.331,-0.476) (0.202,0.007) (-0.071,0.233) (-0.245,0.082) (-0.068,-0.414) (-0.362,-0.042) (0.017,-0.476) (0.223,0.382) (-0.261,0.109) (0.078,-0.276) (-0.060,0.210) (0.017,-0.704) (0.581,-0.267) (0.016,-0.739) (0.398,-0.214) (-0.139,-0.145) (0.224,0.277) (-0.218,0.475) (0.593,0.045) (-0.296,-0.223) (0.018,0.655) (0.064,0.115) (0.091,0.215) (0.117,0.669) (0.129,0.112) (0.472,-0.344) (0.454,-0.069) (0.031,0.548) (0.607,-0.122)};  \path plot[mark=diamond*,mark options={color=black},mark size=1pt] coordinates { (0.384,-0.033) (0.166,0.085) (-0.726,-0.058) (-0.661,0.052) (-0.408,0.069) (-0.353,0.218) (0.040,-0.019) (0.338,0.255) (-0.450,-0.240) (0.535,0.164) (0.503,0.057) (-0.837,-0.119) (0.843,-0.067) (0.462,0.121) (0.465,0.078) (-0.217,0.019) (-0.085,-0.015) (-0.319,-0.160) (0.211,-0.171) (-0.533,-0.128) (-0.278,0.245) (-1.000,0.035) (-0.398,-0.260) (-0.334,-0.238) (-0.379,0.303) (-0.130,0.034) (0.302,-0.201) (-0.073,0.194) (0.817,0.011) (0.184,-0.269)};  \end{tikzpicture}}}
 & 0.296 & \raisebox{-.5\height}{\resizebox {1.2cm} {1.2cm} {\begin{tikzpicture}[scale=6] \path (-1.2,-1.2) (1.2,1.2);\draw[very thin,color=gray] (0,1.1)--(0,-1.1); \draw[very thin,color=gray] (1.1,0)--(-1.1,0); \path plot[mark=square*,mark options={color=red},mark size=0.8pt] coordinates { (-0.345,0.414) (-0.331,-0.476) (0.202,0.007) (-0.071,0.233) (-0.245,0.082) (-0.068,-0.414) (-0.362,-0.042) (0.017,-0.476) (0.223,0.382) (-0.261,0.109) (0.078,-0.276) (-0.060,0.210) (0.017,-0.704) (0.581,-0.267) (0.016,-0.739) (0.398,-0.214) (-0.139,-0.145) (0.224,0.277) (-0.218,0.475) (0.593,0.045) (-0.296,-0.223) (0.018,0.655) (0.064,0.115) (0.091,0.215) (0.117,0.669) (0.129,0.112) (0.472,-0.344) (0.454,-0.069) (0.031,0.548) (0.607,-0.122)};  \path plot[mark=diamond*,mark options={color=black},mark size=1pt] coordinates { (0.384,-0.033) (0.166,0.085) (-0.726,-0.058) (-0.661,0.052) (-0.408,0.069) (-0.353,0.218) (0.040,-0.019) (0.338,0.255) (-0.450,-0.240) (0.535,0.164) (0.503,0.057) (-0.837,-0.119) (0.843,-0.067) (0.462,0.121) (0.465,0.078) (-0.217,0.019) (-0.085,-0.015) (-0.319,-0.160) (0.211,-0.171) (-0.533,-0.128) (-0.278,0.245) (-1.000,0.035) (-0.398,-0.260) (-0.334,-0.238) (-0.379,0.303) (-0.130,0.034) (0.302,-0.201) (-0.073,0.194) (0.817,0.011) (0.184,-0.269)};  \end{tikzpicture}}}
 & 0.296 & \raisebox{-.5\height}{\resizebox {1.2cm} {1.2cm} {\begin{tikzpicture}[scale=6] \path (-1.2,-1.2) (1.2,1.2);\draw[very thin,color=gray] (0,1.1)--(0,-1.1); \draw[very thin,color=gray] (1.1,0)--(-1.1,0); \path plot[mark=square*,mark options={color=red},mark size=0.8pt] coordinates { (-0.345,0.414) (-0.331,-0.476) (0.202,0.007) (-0.071,0.233) (-0.245,0.082) (-0.068,-0.414) (-0.362,-0.042) (0.017,-0.476) (0.223,0.382) (-0.261,0.109) (0.078,-0.276) (-0.060,0.210) (0.017,-0.704) (0.581,-0.267) (0.016,-0.739) (0.398,-0.214) (-0.139,-0.145) (0.224,0.277) (-0.218,0.475) (0.593,0.045) (-0.296,-0.223) (0.018,0.655) (0.064,0.115) (0.091,0.215) (0.117,0.669) (0.129,0.112) (0.472,-0.344) (0.454,-0.069) (0.031,0.548) (0.607,-0.122)};  \path plot[mark=diamond*,mark options={color=black},mark size=1pt] coordinates { (0.384,-0.033) (0.166,0.085) (-0.726,-0.058) (-0.661,0.052) (-0.408,0.069) (-0.353,0.218) (0.040,-0.019) (0.338,0.255) (-0.450,-0.240) (0.535,0.164) (0.503,0.057) (-0.837,-0.119) (0.843,-0.067) (0.462,0.121) (0.465,0.078) (-0.217,0.019) (-0.085,-0.015) (-0.319,-0.160) (0.211,-0.171) (-0.533,-0.128) (-0.278,0.245) (-1.000,0.035) (-0.398,-0.260) (-0.334,-0.238) (-0.379,0.303) (-0.130,0.034) (0.302,-0.201) (-0.073,0.194) (0.817,0.011) (0.184,-0.269)};  \end{tikzpicture}}}
 & 0.296 & \raisebox{-.5\height}{\resizebox {1.2cm} {1.2cm} {\begin{tikzpicture}[scale=6] \path (-1.2,-1.2) (1.2,1.2);\draw[very thin,color=gray] (0,1.1)--(0,-1.1); \draw[very thin,color=gray] (1.1,0)--(-1.1,0); \path plot[mark=square*,mark options={color=red},mark size=0.8pt] coordinates { (0.343,-0.397) (0.329,0.491) (-0.198,-0.019) (0.075,-0.226) (0.252,-0.078) (0.074,0.410) (0.361,0.058) (-0.020,0.459) (-0.219,-0.387) (0.259,-0.101) (-0.080,0.279) (0.063,-0.200) (-0.010,0.706) (-0.570,0.258) (-0.018,0.727) (-0.392,0.209) (0.135,0.129) (-0.221,-0.279) (0.211,-0.469) (-0.582,-0.047) (0.288,0.211) (-0.008,-0.627) (-0.065,-0.123) (-0.088,-0.214) (-0.112,-0.655) (-0.131,-0.126) (-0.466,0.326) (-0.450,0.061) (-0.027,-0.562) (-0.597,0.120)};  \path plot[mark=diamond*,mark options={color=black},mark size=1pt] coordinates { (-0.388,0.035) (-0.192,-0.099) (0.714,0.044) (0.636,-0.044) (0.390,-0.067) (0.359,-0.225) (-0.036,0.032) (-0.342,-0.256) (0.440,0.213) (-0.517,-0.193) (-0.490,-0.048) (0.824,0.103) (-0.811,0.090) (-0.463,-0.142) (-0.450,-0.091) (0.227,0.023) (0.090,-0.000) (0.316,0.183) (-0.208,0.222) (0.516,0.105) (0.278,-0.245) (1.000,-0.025) (0.396,0.259) (0.321,0.211) (0.374,-0.282) (0.149,-0.003) (-0.299,0.216) (0.060,-0.208) (-0.811,-0.008) (-0.219,0.268)};  \end{tikzpicture}}}
\\

\texttt{phi4}
 & \cellcolor{black!10} $<$0.01 & \cellcolor{black!10}\raisebox{-.5\height}{\resizebox {1.2cm} {1.2cm} {\begin{tikzpicture}[scale=6] \path (-1.2,-1.2) (1.2,1.2);\draw[very thin,color=gray] (0,1.1)--(0,-1.1); \draw[very thin,color=gray] (1.1,0)--(-1.1,0); \path plot[mark=square*,mark options={color=red},mark size=0.8pt] coordinates { (-0.543,-0.102) (-0.620,-0.014) (-0.512,-0.004) (-0.522,-0.078) (-0.580,-0.037) (-0.595,-0.004) (-0.600,-0.026) (-0.561,0.012) (-0.502,0.047) (-0.526,0.005) (-0.558,-0.011) (-0.524,-0.052) (-0.612,0.031) (-0.545,0.076) (-0.572,-0.024) (-0.545,0.041) (-0.517,-0.041) (-0.495,-0.038) (-0.536,-0.027) (-0.515,0.054) (-0.554,-0.057) (-0.532,0.011) (-0.489,-0.082) (-0.535,0.005) (-0.533,-0.005) (-0.508,-0.030) (-0.483,0.054) (-0.523,0.057) (-0.461,-0.011) (-0.491,0.018)};  \path plot[mark=diamond*,mark options={color=black},mark size=1pt] coordinates { (0.632,-0.078) (0.639,0.322) (0.337,0.215) (0.788,-0.325) (0.192,-0.180) (0.051,0.385) (0.594,-0.001) (0.367,-0.192) (0.482,-0.210) (0.502,0.339) (0.303,0.226) (0.354,-0.116) (0.670,-0.025) (0.613,0.145) (0.778,0.261) (0.654,-0.273) (0.652,-0.283) (0.377,0.062) (0.686,0.211) (0.765,0.106) (0.544,-0.538) (1.000,-0.053) (0.432,-0.020) (0.385,0.052) (0.577,0.320) (0.700,-0.328) (0.804,-0.251) (0.401,-0.065) (0.576,0.632) (0.236,-0.109)};  \end{tikzpicture}}}
 & \cellcolor{black!10} $<$0.01 & \cellcolor{black!10}\raisebox{-.5\height}{\resizebox {1.2cm} {1.2cm} {\begin{tikzpicture}[scale=6] \path (-1.2,-1.2) (1.2,1.2);\draw[very thin,color=gray] (0,1.1)--(0,-1.1); \draw[very thin,color=gray] (1.1,0)--(-1.1,0); \path plot[mark=square*,mark options={color=red},mark size=0.8pt] coordinates { (-0.573,0.102) (-0.654,0.042) (-0.545,-0.042) (-0.550,0.079) (-0.617,0.033) (-0.628,0.008) (-0.637,0.075) (-0.593,0.021) (-0.528,-0.074) (-0.557,0.010) (-0.588,0.024) (-0.553,0.070) (-0.648,-0.027) (-0.562,-0.087) (-0.598,0.016) (-0.566,-0.055) (-0.552,0.079) (-0.517,0.025) (-0.563,0.053) (-0.528,-0.074) (-0.595,0.072) (-0.556,-0.036) (-0.510,0.101) (-0.560,0.003) (-0.558,-0.004) (-0.534,0.042) (-0.501,-0.095) (-0.545,-0.071) (-0.484,-0.028) (-0.504,-0.048)};  \path plot[mark=diamond*,mark options={color=black},mark size=1pt] coordinates { (0.712,0.147) (0.075,0.092) (0.674,-0.651) (1.000,-0.198) (0.398,0.310) (0.422,-0.091) (0.793,-0.070) (0.444,0.302) (0.413,-0.319) (0.652,-0.270) (0.725,-0.098) (0.997,0.721) (0.446,0.186) (0.588,0.364) (0.754,-0.070) (0.644,0.079) (0.258,-0.354) (0.812,-0.050) (0.565,-0.537) (0.387,0.504) (0.569,0.127) (0.420,-0.260) (0.531,0.119) (0.551,-0.105) (0.336,0.043) (-0.009,-0.275) (0.797,0.226) (0.635,-0.160) (0.665,-0.003) (0.649,0.075)};  \end{tikzpicture}}}
 & \cellcolor{black!10} $<$0.01 & \cellcolor{black!10}\raisebox{-.5\height}{\resizebox {1.2cm} {1.2cm} {\begin{tikzpicture}[scale=6] \path (-1.2,-1.2) (1.2,1.2);\draw[very thin,color=gray] (0,1.1)--(0,-1.1); \draw[very thin,color=gray] (1.1,0)--(-1.1,0); \path plot[mark=square*,mark options={color=red},mark size=0.8pt] coordinates { (-0.573,0.102) (-0.654,0.042) (-0.545,-0.042) (-0.550,0.079) (-0.617,0.033) (-0.628,0.008) (-0.637,0.075) (-0.593,0.021) (-0.528,-0.074) (-0.557,0.010) (-0.588,0.024) (-0.553,0.070) (-0.648,-0.027) (-0.562,-0.087) (-0.598,0.016) (-0.566,-0.055) (-0.552,0.079) (-0.517,0.025) (-0.563,0.053) (-0.528,-0.074) (-0.595,0.072) (-0.556,-0.036) (-0.510,0.101) (-0.560,0.003) (-0.558,-0.004) (-0.534,0.042) (-0.501,-0.095) (-0.545,-0.071) (-0.484,-0.028) (-0.504,-0.048)};  \path plot[mark=diamond*,mark options={color=black},mark size=1pt] coordinates { (0.712,0.147) (0.075,0.092) (0.674,-0.651) (1.000,-0.198) (0.398,0.310) (0.422,-0.091) (0.793,-0.070) (0.444,0.302) (0.413,-0.319) (0.652,-0.270) (0.725,-0.098) (0.997,0.721) (0.446,0.186) (0.588,0.364) (0.754,-0.070) (0.644,0.079) (0.258,-0.354) (0.812,-0.050) (0.565,-0.537) (0.387,0.504) (0.569,0.127) (0.420,-0.260) (0.531,0.119) (0.551,-0.105) (0.336,0.043) (-0.009,-0.275) (0.797,0.226) (0.635,-0.160) (0.665,-0.003) (0.649,0.075)};  \end{tikzpicture}}}
 & \cellcolor{black!10} $<$0.01 & \cellcolor{black!10}\raisebox{-.5\height}{\resizebox {1.2cm} {1.2cm} {\begin{tikzpicture}[scale=6] \path (-1.2,-1.2) (1.2,1.2);\draw[very thin,color=gray] (0,1.1)--(0,-1.1); \draw[very thin,color=gray] (1.1,0)--(-1.1,0); \path plot[mark=square*,mark options={color=red},mark size=0.8pt] coordinates { (-0.573,0.102) (-0.654,0.042) (-0.545,-0.042) (-0.550,0.079) (-0.617,0.033) (-0.628,0.008) (-0.637,0.075) (-0.593,0.021) (-0.528,-0.074) (-0.557,0.010) (-0.588,0.024) (-0.553,0.070) (-0.648,-0.027) (-0.562,-0.087) (-0.598,0.016) (-0.566,-0.055) (-0.552,0.079) (-0.517,0.025) (-0.563,0.053) (-0.528,-0.074) (-0.595,0.072) (-0.556,-0.036) (-0.510,0.101) (-0.560,0.003) (-0.558,-0.004) (-0.534,0.042) (-0.501,-0.095) (-0.545,-0.071) (-0.484,-0.028) (-0.504,-0.048)};  \path plot[mark=diamond*,mark options={color=black},mark size=1pt] coordinates { (0.712,0.147) (0.075,0.092) (0.674,-0.651) (1.000,-0.198) (0.398,0.310) (0.422,-0.091) (0.793,-0.070) (0.444,0.302) (0.413,-0.319) (0.652,-0.270) (0.725,-0.098) (0.997,0.721) (0.446,0.186) (0.588,0.364) (0.754,-0.070) (0.644,0.079) (0.258,-0.354) (0.812,-0.050) (0.565,-0.537) (0.387,0.504) (0.569,0.127) (0.420,-0.260) (0.531,0.119) (0.551,-0.105) (0.336,0.043) (-0.009,-0.275) (0.797,0.226) (0.635,-0.160) (0.665,-0.003) (0.649,0.075)};  \end{tikzpicture}}}
 & \cellcolor{black!10} $<$0.01 & \cellcolor{black!10}\raisebox{-.5\height}{\resizebox {1.2cm} {1.2cm} {\begin{tikzpicture}[scale=6] \path (-1.2,-1.2) (1.2,1.2);\draw[very thin,color=gray] (0,1.1)--(0,-1.1); \draw[very thin,color=gray] (1.1,0)--(-1.1,0); \path plot[mark=square*,mark options={color=red},mark size=0.8pt] coordinates { (-0.543,-0.102) (-0.620,-0.014) (-0.512,-0.004) (-0.522,-0.078) (-0.580,-0.037) (-0.595,-0.004) (-0.600,-0.026) (-0.561,0.012) (-0.502,0.047) (-0.526,0.005) (-0.558,-0.011) (-0.524,-0.052) (-0.612,0.031) (-0.545,0.076) (-0.572,-0.024) (-0.545,0.041) (-0.517,-0.041) (-0.495,-0.038) (-0.536,-0.027) (-0.515,0.054) (-0.554,-0.057) (-0.532,0.011) (-0.489,-0.082) (-0.535,0.005) (-0.533,-0.005) (-0.508,-0.030) (-0.483,0.054) (-0.523,0.057) (-0.461,-0.011) (-0.491,0.018)};  \path plot[mark=diamond*,mark options={color=black},mark size=1pt] coordinates { (0.632,-0.078) (0.639,0.322) (0.337,0.215) (0.788,-0.325) (0.192,-0.180) (0.051,0.385) (0.594,-0.001) (0.367,-0.192) (0.482,-0.210) (0.502,0.339) (0.303,0.226) (0.354,-0.116) (0.670,-0.025) (0.613,0.145) (0.778,0.261) (0.654,-0.273) (0.652,-0.283) (0.377,0.062) (0.686,0.211) (0.765,0.106) (0.544,-0.538) (1.000,-0.053) (0.432,-0.020) (0.385,0.052) (0.577,0.320) (0.700,-0.328) (0.804,-0.251) (0.401,-0.065) (0.576,0.632) (0.236,-0.109)};  \end{tikzpicture}}}
 & \multicolumn{1}{c}{$\times$} & \multicolumn{1}{c}{$\times$}\\

        \bottomrule
    \end{tabular}
    }
    
\end{sidewaystable}

\FloatBarrier

 \bibliographystyle{elsarticle-num}

\end{document}